\documentclass[prc,showpacs,showkeys,superscriptaddress,nofootinbib,twocolumn,floatfix]{revtex4}
\usepackage{graphicx,color,mathrsfs,amsmath,amssymb,amsthm,amsopn,mathbbol,bm} 
\usepackage[nottoc]{tocbibind}
\newlength{\lslashl}

\usepackage[normalem]{ulem} 
\renewcommand\sout{\bgroup \color{red} \ULdepth=-.5ex \ULset}

\begin{document}

\title{Quarkonia and Heavy-Quark Relaxation Times 
in the Quark-Gluon Plasma} 

\author{F. Riek}\thanks{e-mail: friek@comp.tamu.edu}
\affiliation{Cyclotron Institute, Texas A\&M University, 
College Station, Texas 77843-3366, USA}
\author{R. Rapp}\thanks{e-mail: rapp@comp.tamu.edu}
\affiliation{Cyclotron Institute, Texas A\&M University, 
College Station, Texas 77843-3366, USA}

\begin{abstract}
A thermodynamic $T$-matrix approach for elastic 2-body interactions is 
employed to calculate spectral functions of open and hidden heavy-quark 
systems in the Quark-Gluon Plasma. This enables the evaluation of 
quarkonium bound-state properties and heavy-quark diffusion on a common 
basis and thus to obtain mutual constraints. The two-body interaction 
kernel is approximated within a potential picture for spacelike 
momentum transfers.
An effective field-theoretical model combining color-Coulomb and 
confining terms is implemented with relativistic corrections and for 
different color channels.
Four pertinent model parameters, characterizing the coupling strengths
and screening, are adjusted to reproduce the color-average 
heavy-quark free energy as computed in thermal lattice QCD. The 
approach is tested against vacuum spectroscopy in the open ($D$, $B$) 
and hidden ($\Psi$ and $\Upsilon$) flavor sectors, as well as in the 
high-energy limit of elastic perturbative QCD scattering. 
Theoretical uncertainties in the static reduction scheme of the 
4-dimensional Bethe-Salpeter equation are elucidated.  
The quarkonium spectral functions are used to calculate Euclidean 
correlators which are discussed in light of lattice QCD results,
while heavy-quark relaxation rates and diffusion coefficients are 
extracted utilizing a Fokker-Planck equation. 
\end{abstract}

\date{\today}
\maketitle

%
\section{Introduction}
\label{sec_intro}
%
Hadrons containing heavy quarks are widely used to deduce basic 
properties of the strong force as given by Quantum Chromodynamics 
(QCD)~\cite{Brambilla:2004wf}. Heavy-quark (HQ) systems are also 
valuable for studying hot and dense matter, in particular for 
temperatures and quark chemical potentials which are parametrically 
small compared to the HQ mass, $T , \mu_q \ll m_Q$ ($Q=c,b$). Under 
these conditions, which are believed to encompass phase changes of 
the medium, HQ momenta ($p_Q^2\sim m_Q T$) are large relative to those 
of the (light) partons ($p^2 \sim T^2$) constituting the heat bath. 
This leads to simplifications in the theoretical description which 
facilitate the task of studying the medium. For example, the 
prevalence of elastic interactions with spacelike momentum transfers
suggest that a potential picture for HQ interactions - successful 
in vacuum - may remain valid in the medium.  
   
The modifications of heavy quarkonia (charmonium and bottomonium) 
in the medium are believed to reveal quark deconfinement in the 
Quark-Gluon Plasma (QGP) as produced in ultrarelativistic 
heavy-ion collisions (URHICs), 
cf.~Refs.~\cite{Rapp:2008tf,BraunMunzinger:2009ih,Kluberg:2009wc} 
for recent reviews. In addition, open heavy-flavor particles are being
utilized to extract transport properties of the medium formed at the 
Relativistic Heavy-Ion Collider (RHIC), by computing their diffusion 
coefficient in the QGP, see, e.g., Ref.~\cite{Rapp:2009my} for a recent 
review. In Ref.~\cite{vanHees:2007me} it was suggested that the (in-medium) 
forces responsible for heavy-quarkonium binding may be closely related to 
those governing the diffusion of an individual heavy quark in the QGP. 
The basis for this idea is that the exchanged 4-momentum, 
$k=(k_0,\vec k\,)$, for the elastic scattering of a (slow) heavy quark 
is essentially ``static", i.e., the energy transfer
is parametrically smaller than the 3-momentum transfer, 
$k_0\simeq \vec{k}^{\,2}/2m_Q \ll |\vec k\,|$, for both an individual 
heavy quark and a quarkonium bound-state.
A possible consequence of such a connection could be that a ``strong" 
interaction in the medium, which binds charmonium states above the 
critical temperature, at the same time leads to strong correlations 
in the heavy-light sector, which accelerate HQ thermalization compared 
to perturbative scattering~\cite{vanHees:2007me}.   
The purpose of the present paper is to set up and carry out a framework
which enables a systematic investigation of this connection. This 
requires to evaluate in-medium bound and scattering states on equal 
footing, which will be realized by employing a thermodynamic $T$-matrix 
approach~\cite{Mannarelli:2005pz,Cabrera:2006wh}.  
To improve the reliability of this framework, several constraints
will be elaborated: input potentials will be adjusted to reproduce the
HQ free energy computed in lattice QCD (lQCD) in vacuum and at finite 
temperature, empirical vacuum spectroscopy in the bound-state regime 
and perturbative QCD in the high-energy scattering limit will be 
checked, and euclidean correlators for heavy quarkonia in medium 
will be tested with lQCD results.   

In the vacuum, the description of heavy quarkonia within a potential
framework can be made rigorous as an effective field theory of QCD with 
heavy quarks, so-called potential non-relativistic QCD 
(pNRQCD)~\cite{Brambilla:2004wf}. In a hot and dense medium, however,
additional scales enter the problem (e.g., temperature, $T$, and Debye 
screening mass, $m_D$) rendering the extension of the potential concept 
more involved, especially in a strongly interacting system where it is 
difficult to establish scale hierarchies (for perturbative treatments 
based on $T\gg m_D \sim gT$, see, e.g., 
Refs.~\cite{Laine:2006ns,Beraudo:2007ky,Brambilla:2008cx}).
On the other hand, nonperturbative information on the HQ interaction
over a wide range of temperatures is available from thermal lQCD in 
terms of the free energy, $F$, of a static $Q\bar Q$ pair (strictly 
speaking, the difference in free energy of the system with and without 
the HQ pair)~\cite{Petreczky:2004pz,Kaczmarek:2005ui,Kaczmarek:2007pb}.
In practical approaches, the color-singlet free energy, $F_1$ (or 
the pertinent internal energy, $U_1=F_1-T\,\partial F_1/\partial T$) 
has been utilized as a potential in
Schr\"odinger~\cite{Mocsy:2005qw,Wong:2006bx,Alberico:2006vw,Dumitru:2009ni} 
and $T$-matrix~\cite{Cabrera:2006wh,Mannarelli:2005pz} equations, and the resulting
spectral functions have been checked against lQCD results for euclidean 
correlation functions. While these approaches suggest that the potential
model provides a viable tool at finite temperature, several open issues 
remain, e.g.: 
(i) the use of free or internal energy (or even 
combinations thereof~\cite{Wong:2004zr}); 
(ii) the gauge dependence of the color-singlet free 
energy~\cite{Philipsen:2008qx}; 
(iii) microscopic insights into the screening mechanisms (e.g., 
color-Coulomb vs. confining forces);
(iv) corrections to (or even validity of) the potential ansatz.  
In the present paper we do not offer new principle insights on item (i). 
To cover the uncertainty associated with this problem, our calculations 
will be carried out for both free and internal energies which are 
believed to bracket the limiting cases within their interpretation as 
a finite-temperature HQ potential.  
To address items (ii) and (iii) we adopt a recently proposed 
field-theoretic ansatz~\cite{Megias:2007pq,Megias:2005ve} to describe 
HQ free energies using a screened Coulomb plus ``confining" gluon 
propagator. These propagators require four input parameters (coupling 
strengths and screening masses) which are adjusted to reproduce
gauge-invariant color-average free energies from lQCD. 
Color projections are extracted within the model and utilized to 
compute color-singlet quarkonium and heavy-light spectral functions, 
as well as colored correlations which contribute to the transport  
of heavy quarks in the QGP.
Special care is taken of relativistic effects -- 
especially for light quarks -- which in our framework is possible once 
the vector and/or scalar nature of the Coulomb and confining force is 
specified. For example, the relativistic Breit correction known from 
electrodynamics~\cite{Brown:1952} naturally emerges as a relativistic
effect in the Coulomb potential. We will 
furthermore check the static approximation underlying the potential 
picture, by comparing different versions of the 3-dimensional reduction 
scheme to obtain the $T$-matrix from an underlying Bethe-Salpeter 
equation. 

This article is organized as follows. In Sec.~\ref{sec_megias} we set 
up the microscopic model used to fit lQCD free energies. In 
Sec.~\ref{sec_tmat} we recollect main elements of the thermodynamic 
$T$-matrix formalism, including relativistic corrections. 
Section~\ref{sec_sf} is devoted to the discussion of our numerical 
results for quarkonium and heavy-light spectral functions and their 
applications to euclidean correlators and HQ relaxation times, 
respectively. We conclude in Sec.~\ref{sec_concl}.
 
%
\section{Microscopic Model for the Heavy-Quark Potential} 
\label{sec_megias}
The recent revival of potential models to describe heavy quarkonia in medium
has been largely driven by the prospect of a paremeter-free input via static 
HQ free energies computed in thermal lQCD.
However, functional fits to the lattice ``data" usually do not offer much 
insight into the physical mechanisms underlying the medium effects in the 
potential, nor do they allow to define vertex structures of the interaction 
which become important at higher energies and/or in different color channels. 
In addition, it is desirable to base the starting point on a gauge-invariant 
quantity, i.e., the color-average free energy\footnote{Since the model 
adopted here and the lattice calculations use different gauges (static vs. 
Coulomb), this is the only meaningful way to extract parameters.}. 
To this end, we adopt the microscopic model developed by Meg\'{i}as et 
al.~\cite{Megias:2007pq,Megias:2005ve} which we briefly review in the 
following and then fit to lattice data. 
The key idea underlying this model 
is that the HQ free energy can be accounted for by a nonperturbative ansatz
for the gluon propagator giving rise to a string-like confining term in
coordinate space, plus a ``standard" perturbative term corresponding 
to a screened color-Coulomb interaction~\cite{Megias:2007pq}, 
\begin{eqnarray}
&&D_{00}(\vec{k}\,)=D^P_{00}(\vec{k}\,)+D^{NP}_{00}(\vec{k}\,) 
\label{D00}
\\
&&D^P_{00}(\vec{k}\,)=\frac{1}{{\vec k}^2+m_D^2} \qquad 
D^{NP}_{00}(\vec{k}\,)=\frac{m_{G}^2}{({\vec k}^2+\tilde{m}_{D}^2)^2}
\nonumber
\end{eqnarray}
to be understood in static gauge and within dimensional reduction
($m_D$, $\tilde{m}_D$: screening masses). The leading nonperturbative 
effect in the gluon propagator is associated with a dimension-2 gluon 
condensate dictated by dimensional considerations, 
\begin{equation}
\langle A^2_{0,a}\rangle^{NP}= T\,\frac{(N_c^2-1) m_G^2}{8\pi\tilde{m}_D} \ , 
\end{equation}
with $m_G^2$ a ``glueball" mass. 
A priori, a dimension-2 gluon condensate is gauge-dependent and as such 
a somewhat controversial quantity. Since the gluon propagator is a 
gauge-dependent quantity the appearance of gauge-dependent terms
is inevitable. However, it has been argued by several 
authors~\cite{Lavelle:1988eg,Gubarev:2000nz,Gubarev:2000eu,Kondo:2001nq,Dudal:2009ti}
that a dimension-2 condensate encodes nontrivial gauge-invariant 
information, e.g., topological configurations associated with magnetic 
monopoles giving rise to a static confining force (which is precisely the 
effect modeled in the present context). Evidence for a dimension-2  
condensate contribution has also been found in QCD sumrules (see 
Ref.~\cite{Boucaud:2000nd} and references therein).

To establish the connection to the HQ free energy (given by a correlator
of Polyakov loops, $\Omega$), one starts from its perturbative expression
at finite temperature in the color-singlet channel~\cite{Megias:2007pq},
\begin{eqnarray}
&&{\rm e}^{-F_1(r,T)/T}=
\langle\frac{1}{N_c}Tr\left[\Omega(x)\Omega^\dag(y)\right]\rangle
\nonumber\\
&&={\rm e}^{(g^2/(2N_cT^2))\,\langle A_{0,a}(x)A_{0,a}(y)-A^2_{0,a}(x)\rangle}
+\mathcal{O}(g^6)
\label{F1_pert}
\end{eqnarray}
with the perturbative gluon propagator 
\begin{eqnarray}
\langle A_{0,a}(x)A_{0,b}(y)\rangle = \delta_{ab} T
\int \frac{d^3k}{(2\pi)^3}e^{i\vec{k}\cdot(\vec{x}-\vec{y}\,)}
D^P_{00}(\vec{k}\,) \ . 
\label{gluon-prop}
\end{eqnarray}
The separation-independent term, $\langle A^2_{0,a}(x)\rangle$, in 
Eq.~(\ref{F1_pert}) plays the role of a selfenergy of an individual heavy 
quark. The main assumption consists now of augmenting the perturbative 
propagator by the nonperturbative
one as given by Eq.~(\ref{D00}). Assuming further that the same functional 
dependence as in Eq.~(\ref{F1_pert}) holds in the nonperturbative case we 
are lead to the following form of the singlet free energy ($N_c=3$):
\begin{eqnarray}
F_1(r,T)
&=&-\frac{4}{3}\alpha_s\left(\frac{1}{r}{\rm e}^{-m_{D}r}+
\frac{m_{G}^2}{2\tilde{m}_{D}}{\rm e}^{-\tilde{m}_{D}r}\right.
\nonumber\\
&&\qquad\quad -\left. \frac{m_{G}^2}{2\tilde{m}_{D}}+ m_{D}\right) \ .
\label{F1_np}
\end{eqnarray}
In Ref.~\cite{Megias:2007pq} this approach has been applied to study the 
Wilson loop and HQ free energy in quenched QCD at finite temperature, and
efficiently reproduces pertinent lQCD data. As an extension of this 
treatment we allow for different screening masses in the perturbative and 
nonperturbative parts of the gluon propagator which improves the 
precision in our fits to unquenched lQCD data. As indicated above we fit 
the color-average free energy. Since we aim at a parametrization over a 
large range in distance and temperature we employ the 
definition~\cite{Petreczky:2005bd}
\begin{eqnarray}
F_{av}(r,T)=
-T\,\ln\left[\frac{1}{9}e^{-F_1(r,T)/T}+\frac{8}{9}e^{-F_8(r,T)/T}\right]
\label{Fav-def}
\end{eqnarray}
without further approximations which automatically ensures the correct 
behavior in the limits $rT>>1$, where the potential is dominated by 2-gluon 
exchange, $F_{av.}(r,T)\approx (F_1(r,T))^2$, and $rT<<1$, where we have 
$F_{av}(r,T)\approx F_1(r,T)+T\ln(9)$~\cite{Kaczmarek:2002mc}. The 
evaluation of Eq.~(\ref{Fav-def}) requires the color-octet and -singlet
free energies (in addition we will use the extracted potentials in the sextet 
and anti-triplet channels for the calculation of HQ relaxation times). In 
previous works~\cite{Mannarelli:2005pz,vanHees:2007me} these potentials have 
been approximated by Casimir scaling of the singlet potential. While this is 
a good approximation for the short-range (perturbative/Coulombic) part of the 
potential~\cite{Kaczmarek:2002mc,Doring:2007uh}, it presumably does not 
apply to the confining part which rather appears to be universal, 
including its long-distance 
limit~\cite{Digal:2003jc,Zantow:2003ui,Petreczky:2005bd}\footnote{Calculations
reported in Ref.~\cite{Nakamura:2005hk} come to a different conclusion, 
possibly because they cover smaller distances than those in 
Refs.~\cite{Zantow:2003ui,Petreczky:2005bd}. In Ref.~\cite{Petreczky:2005bd} 
potential problems with the computation of the octet free energy on the 
lattice below $T_c$ have been pointed out.}. 
We therefore apply Casimir scaling only to the Coulomb part of the model
gluon propagator while the string part is assumed to be color-blind, i.e., 
the same in all color channels\footnote{Such a decomposition is not possible 
in functional parametrizations of the potential.}.  
\begin{figure*}[t]
\includegraphics[scale=1.0]{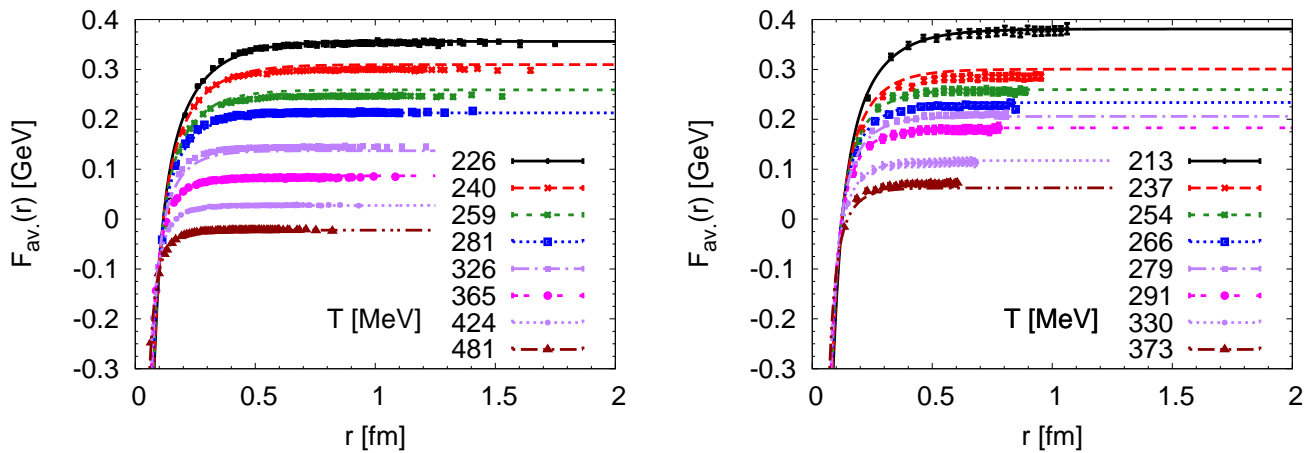}
\caption{(Color online) Color-average HQ free energies obtained in thermal
lQCD computations by Kaczmarek et al. (left
panel)~\cite{Kaczmarek:2007pb,Kaczmarek-private} and Petreczky et al.~(right
panel)~\cite{Petreczky:2004pz,Petreczky-private}, compared to our fits using
the microscopic model suggested in Refs.~\cite{Megias:2007pq,Megias:2005ve}.
}
\label{Fav}
\end{figure*}
\begin{figure*}[t]
\includegraphics[scale=1.0]{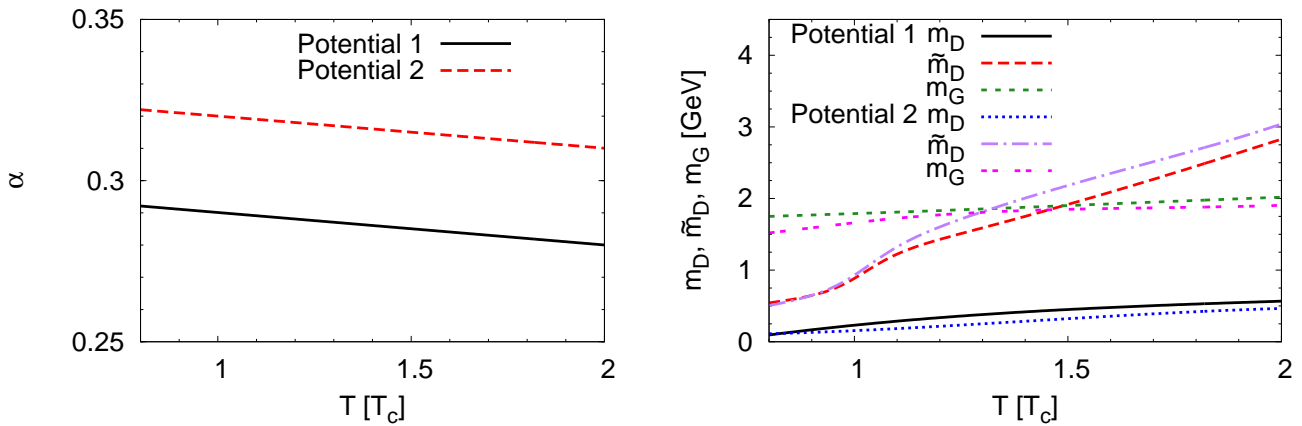}
\caption{(Color online) Temperature dependence of the parameters deduced 
from our fit to the color-average free energies from the two lQCD 
calculations displayed in Fig.~\ref{Fav}. 
Left panel: strong coupling constant $\alpha_s$; Right panel: 
screening masses of the color-Coulomb and confining parts, and dimension-2
condensate ``glueball mass", $m_G$.}
\label{fig_para}
\end{figure*}
This is also compatible with the interpretation of the long-distance
limit as an individual HQ selfenergy, as discussed below.
The coordinate-space potential in a color-channel $a$ takes the form 
\begin{eqnarray}
F_a(r,T)
&=&-\frac{4}{3}\alpha_s\left(\frac{C_a}{r}e^{-m_{D}r}
  +\frac{m_{G}^2}{2\tilde{m}_{D}}e^{-\tilde{m}_{D}r} \right.
\nonumber \\
&& \qquad \quad \ \ \left. -\frac{m_{G}^2}{2\tilde{m}_{D}}+m_{D}\right) 
\nonumber \\
&\equiv&F_a^{C}(r,T)+F^{S}(r,T)+ F_\infty(T)
\label{Fcolor-Meg}
\\
C_1=1 && C_8=-1/8 \qquad  C_6=-1/4 \qquad  C_{\bar{3}}=1/2 \ ,
\nonumber 
\end{eqnarray}
with the Coulomb part $F_a^{C}$, the nonperturbative string part $F^{S}$, 
and a $r$-independent part, $F_\infty(T)$,  which will be associated 
with a (real part of the) HQ ``selfenergy", $\Sigma_Q^R$, below. 
Similar analytic forms of the potential have been used for fits to the 
color-singlet potential in Ref.~\cite{Dumitru:2009ni}. Let us examine two 
limits of this expression. First, for $T\,\rightarrow\,0$ both screening 
masses should vanish while the condensate characterized by $m_G$ remains 
finite~\cite{Megias:2007pq};
one obtains~\cite{Megias:2007pq}
\begin{eqnarray}
F_a(r,T=0)=
-\frac{4}{3}\alpha_s\frac{C_a}{r}+\sigma\,r\quad , 
\quad \sigma=\frac{2\,\alpha_s\,m_{G}^2}{3}
\label{Megias-Fa-T0}
\end{eqnarray}
which recovers the Cornell potential in the color-singlet channel and 
yields a universal string tension in all color channels, consistent 
with Refs.~\cite{Digal:2003jc,Zantow:2003ui,Petreczky:2005bd}. Second, 
for $r\to\infty$ at finite $T$ one has 
\begin{eqnarray}
F_\infty \equiv F_a(r\rightarrow\infty,T)
=\frac{4}{3}\alpha_s\left(-m_{D}+\frac{m_{G}^2}{2\tilde{m}_{D}}\right)
\label{F-rinf}
\end{eqnarray}
which is independent of $a$, consistent with lQCD data in 
Refs.~\cite{Kaczmarek:2005ui,Petreczky:2005bd}.
When additionally taking the zero-temperature limit of $F_\infty$, 
it diverges since $m_G>0$ and $\tilde{m}_{D}\,\rightarrow\,0$. This is,
of course, expected in quenched QCD but needs to be amended in the
presence of light quarks, to simulate ``string breaking". 
We implement this by enforcing a flat potential above a string-breaking scale 
of about $r\simeq 1.1$~fm where the potential has reached about 1.1\,GeV. 
We now also see that the cancellation between the leading piece of the second 
term and the constant third term in the parentheses of Eq.~(\ref{Fcolor-Meg}) 
in the $\tilde{m}_{D}\,\rightarrow\,0$ limit only works for all color channels 
if the string term is color-blind (i.e., has no Casimir scaling).    
If, on the other hand, both terms proportional to $m_G^2$ are subject
to Casimir scaling, it would imply that the $r\to\infty$ limit (i.e.,
the single HQ selfenergy) picks up a strong dependence on the color 
orientation of the quark, which is not natural.

Once the temperature dependence of the parameters $m_G$, $m_D$ and 
$\tilde{m}_D$, as well as of the strong coupling, $\alpha_s$, is determined
through fits of the color-average free energy, corresponding expressions 
for the internal energy, $U$, can be computed via 
\begin{equation}
U(r,T)=F(r,T)-T\frac{d}{dT}F(r,T)  
\label{U}
\end{equation}
and projected into different color-channels, $a$.
\begin{figure*}[t]
\includegraphics[scale=1.0]{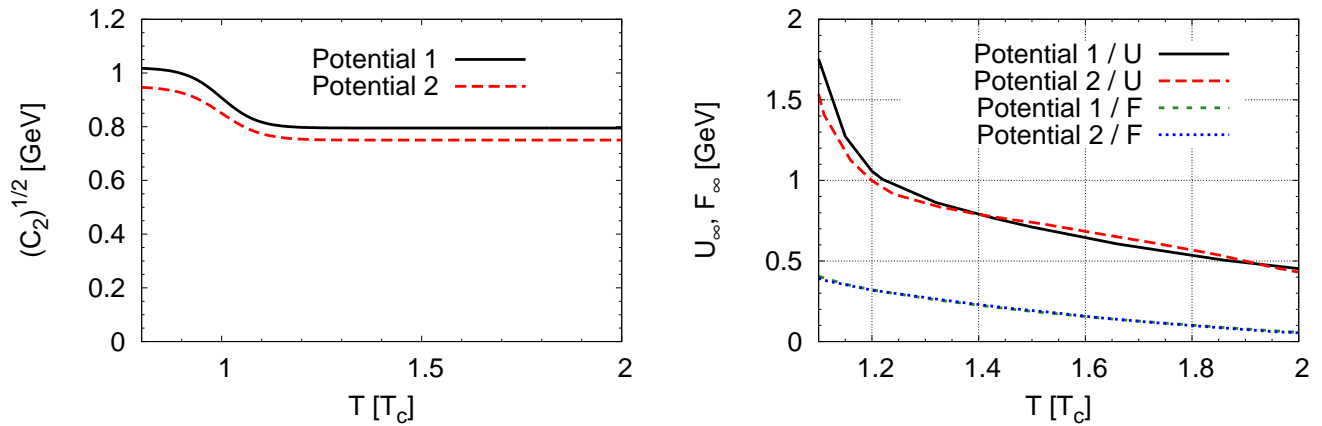}
\caption{(Color online) Temperature dependence of the dimension-2 condensate 
$C_2$, Eq.~(\ref{dim-two-condensate}), (left panel) and of the 
infinite-distance limit of the free and internal energies for the two lQCD 
computations displayed in Fig.~\ref{Fav}.} 
\label{fig_C2-Vinf}
\end{figure*}
\begin{figure*}[t]
\includegraphics[scale=1.0]{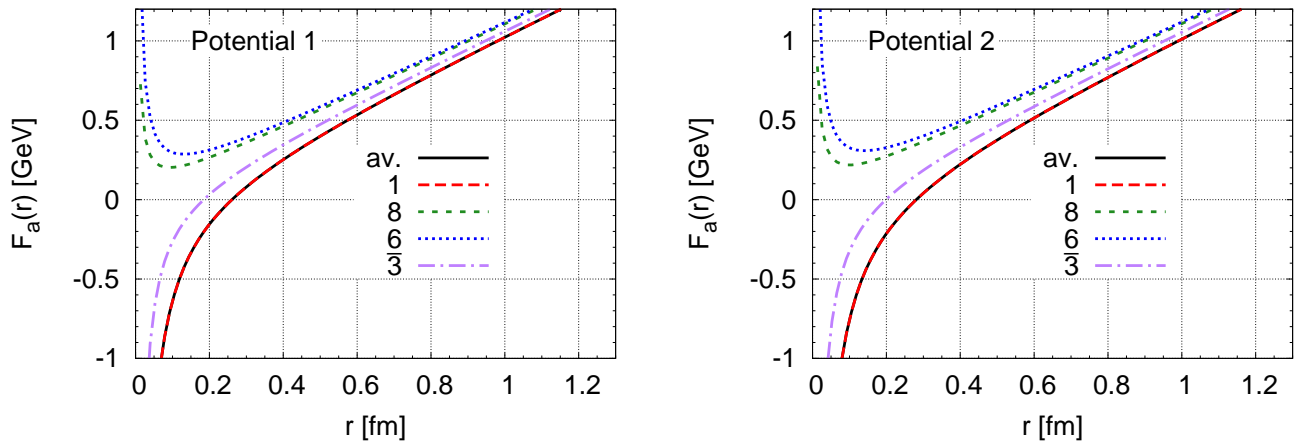}
\caption{(Color online) Vacuum HQ potentials in different color configurations
of $Q\bar Q$ (singlet and octet) and $QQ$ (triplet and sextet) channels,
extracted from model-fits to color-average $Q\bar Q$ free energies in lQCD
for $N_f=2+1$~\cite{Kaczmarek:2007pb,Cheng:2007jq,Kaczmarek-private} (``potential
 1", left panel) and for $N_f=3$~\cite{Petreczky:2004pz,Petreczky-private}
(``potential 2", right panel).}
\label{fig_VT0}
\end{figure*}

It is currently an unsettled question whether the free or internal energy 
(or a linear combination thereof) is a more suitable quantity to be utilized 
as a static two-body potential in a Schr\"odinger and/or scattering equation. 
We recall that the quantity, $F(r,T)$, computed in thermal lQCD, is the 
difference between the free energies of the thermal system containing a 
static $Q\bar Q$ pair and the thermal system without the pair. In 
Ref.~\cite{Satz:2008zc} it has been argued that the pertinent difference 
in internal energies, $U(r,T)$, recovers the thermal expectation value of 
the potential energy between the static $Q$ and $\bar Q$ charges. This 
suggests $U(r)$ as the appropriate in-medium two-body potential. Such a 
potential would by construction be a real
quantity and thus a natural starting point to be unitarized in a scattering
equation, generating appropriate on-shell cuts in the intermediate state 
through imaginary parts in the scattering amplitude.   
In Ref.~\cite{Shuryak:2004tx} it has been argued that the relevance of $F$ vs. 
$U$ depends on the interplay of the thermal relaxation time in the heat bath 
and the interaction time of the $Q$ and $\bar Q$. If the former is much shorter 
than the latter, the $Q\bar Q$ motion will be adiabatic and the free energy 
should be used; on the other hand, for very short interaction times (e.g., for
a broad resonance or high-energy scattering), the internal energy should be
more suitable.
Another approach to the problem has been pursued by using effective field 
theories at finite temperature~\cite{Beraudo:2007ky,Brambilla:2008cx},
by combining HQ and perturbative hierarchies (HQ speed $v\ll1$ and $gT\ll T$). 
These studies recover the color-Coulomb part of the interaction and suggest 
that the free energy plays the role of a potential in a Schr\"odinger equation. 
In addition, an imaginary part of the effective potential has been 
identified~\cite{Laine:2006ns,Rothkopf:2009pk}. 
We emphasize, however, that within a $T$-matrix approach, imaginary 
parts are generated via the unitarization procedure in the intermediate 
propagator and thus imaginary parts in the potential should be 
implemented via suitable cuts in a coupled-channel treatment.
To account for the present uncertainty in the identification of an irreducible 
2-body HQ potential we will show numerical results for both $F$ and $U$ as
driving kernel in the $T$-matrix equation. More precisely, we utilize
the subtracted quantity 
\begin{equation}
V_a(r;T)=X_a(r,T)-X(r\to\infty,T) \ , \ \ X=F \ {\rm or} \ U
\label{V}
\end{equation}
to ensure convergence of the Fourier transform. These choices are 
believed to bracket the range of interaction strength in the HQ sector.
\begin{figure*}[t]
\includegraphics[scale=0.7]{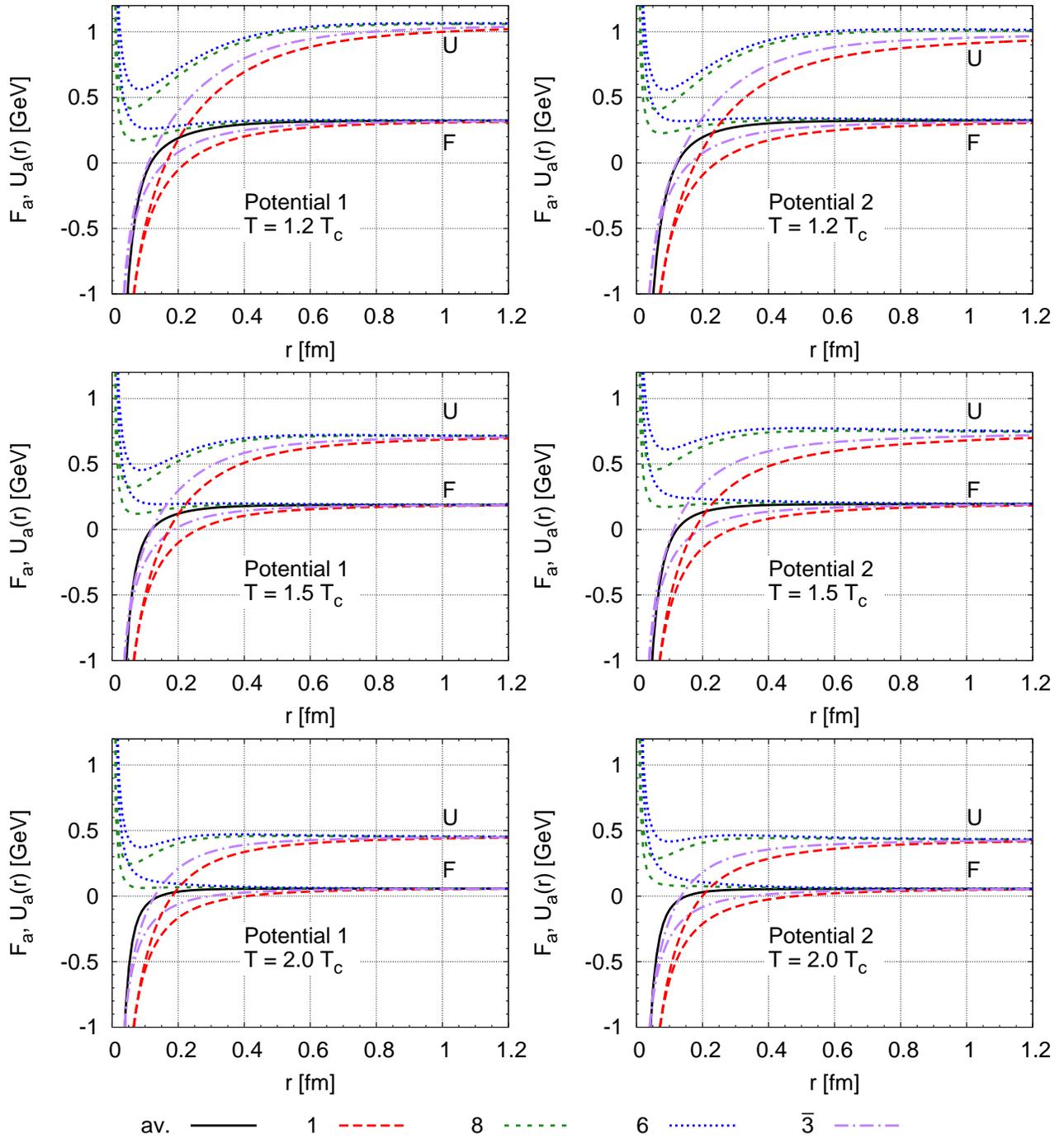}
\caption{(Color online) In-medium heavy-quark free and internal energies
at various temperatures in color-singlet (long-dashed lines), -octet 
(short-dashed lines), -anti-triplet (dash-dotted lines), -sextet (dotted lines)
and -average (solid lines) channels, extracted from $N_f$=2+1 (left column) 
and $N_f$=3 lQCD computations (right column).}
\label{fig_VT}
\end{figure*}

Besides the potential, another important ingredient to the two-body scattering
equation are the in-medium selfenergies of the heavy quark and antiquark,
which we treat symmetrically as appropriate for a hot medium at vanishing
baryon chemical potential. Starting from a bare quark of mass $m_Q^0$, we 
interpret the potential value at infinite distance, $X(r\to\infty,T)$ in 
Eq.~(\ref{V}), as a temperature-dependent ``mean-field" 
contribution to the HQ masses, i.e., as a real part of the selfenergy,  
\begin{equation}
m_Q=m_Q^0+\Sigma_Q^R(T) \ \ , \ \ \Sigma_Q^R(T)\equiv X(r\to\infty,T)/2 \ .
\label{m-eff}
\end{equation}
This interpretation follows from the picture that at infinite distance the 
quarks have become independent from each other which is supported 
by lQCD results as discussed above.
In addition, we will investigate the effects 
of imaginary parts of the HQ selfenergy, 
\begin{equation}
\Gamma_{Q} = -2 \ {\rm Im}\,\Sigma_Q \ ,
\end{equation}
associated with HQ rescattering in the heat bath. 
This quantity has been estimated to be about $\sim$0.2\,GeV in the 
$T$-matrix calculations of Ref.~\cite{vanHees:2007me}.

Let us now turn to the fit of the potentials to recent lQCD data for 
the color-average free energy. To obtain an indication of the systematic 
uncertainty underlying different lQCD inputs we will carry out all our 
calculations for $N_f$=2+1-flavor 
QCD~\cite{Kaczmarek:2007pb,Cheng:2007jq,Kaczmarek-private} and for
$N_f$=3-flavor QCD~\cite{Petreczky:2004pz,Petreczky-private}(the latter 
input has been used in the color-singlet channel in a previous $T$-matrix 
study~\cite{Cabrera:2006wh}), which we refer to in the following as 
potential 1 and 2, respectively. The underlying (pseudo-) critical
temperatures in these lQCD calculations have been quoted as $T_c$=196\,MeV
(potential 1) and 190\,MeV (potential 2). 
Fig.~\ref{Fav} shows the pertinent color-average free energies together 
with our fits which have been performed down to temperatures of 
ca.~0.8\,$T_c$ (not all are shown in the plot). 
The agreement with each data set is fair and supports the 
adequacy of the underlying model. The temperature dependence of the four 
fit parameters is displayed in Fig.~\ref{fig_para}. The variation of
all parameters between the two potentials is rather small. The strong
coupling constant, $\alpha_s(T)$, depends weakly on $T$. To suppress 
fluctuations in an unconstrained fit, we have for simplicity adopted a 
linear ansatz\footnote{Without this constraint, the fits tend to generate 
what we believe are artificially large variations in the parameters,
mainly caused by varying ranges in $r$ as covered by the lQCD data at 
different temperatures. This is particularly evident when fitting 
to the color-singlet free energy and removing some of the small-distance 
points. As a general guiding principle in the fits we tried to utilize 
redundancies in parameter choices to obtain smooth variations with $T$.}.
The screening masses exhibit an appreciable increase with temperature 
reminiscent of the linear $T$-dependence one expects from leading-order 
perturbation theory, $m_D^{\rm pert}= (1+N_f/6)^{1/2} gT$. Compared to
the perturbative result, the coefficients in our fits are smaller for the 
Coulomb part ($m_D$) and larger for the ``string" part ($\tilde m_D$), 
suggesting a stronger screening of the confining part (which primarily 
acts at larger distances).   
We have tried fits enforcing the condition $m_D=\tilde{m}_D$ but could not 
obtain satisfactory accuracy without introducing unnaturally large variations 
of the parameters. This might support the assertion of differences in
the screening process for the perturbative and nonperturbative components
of the free energy. On the other hand, the variation of the glueball mass 
with $T$ is weak. Recalling the explicit relation to the dimension-2 
condensate~\cite{Megias:2007pq} 
\begin{equation}
C_2=g^2\langle A^2_{0,a}\rangle^{NP}=4\alpha_s\, 
T\,\frac{m_G^2}{\tilde{m}_D} \ ,
\label{dim-two-condensate}
\end{equation}
we find that a constant of about $C_2\simeq0.8$\,GeV$^2$ above $T_c$ (see 
left panel of Fig.~\ref{fig_C2-Vinf}) is well compatible with our fit, 
as also found previously in Ref.~\cite{Megias:2007pq} and in analyses of 
the gluon propagator, three-gluon vertex or quark propagator (see 
Ref.~\cite{Megias:2005ve} and references therein). On the other hand, we 
have verified that the 20\% decrease of $C_2(T)$ across $T_c$ is a robust 
feature within ``reasonable"  variations of the other fit parameters; 
e.g., when imposing an overall $T$-independent value for $C_2$, we  
could not reproduce the lQCD values of $F_{av}^{\infty}$ over the entire 
temperature range without ``unnaturally" large variations in the other 
fit parameters. It is tempting to speculate that the ~20\% drop in 
$C_2(T)$ across $T_c$ is related to a similar drop found in the 
magnetic-monopole density in $SU$(2) gluodynamics in 
Ref.~\cite{Gubarev:2000nz}, where qualitative arguments have been put 
forward that an $A_\mu^2$ condensate could be connected to the monopole 
density. The temperature window over which the variation of $C_2(T)$ 
occurs basically conicides with where rapid changes in the 
infinite-distance values $F_\infty$ and $U_\infty$ are observed, 
cf.~right panel of Fig.~\ref{fig_C2-Vinf}. 
In the zero-temperature limit, assuming that both screening masses go to 
zero, a value of $m_G\,\approx\,1$\,GeV is needed to 
reproduce the vacuum string tension of $\sqrt{\sigma}=0.465$\,GeV found in 
lQCD~\cite{Kaczmarek-private,Petreczky-private}, in connection with
strong couplings of $\alpha_s$=0.285 and 0.33 for potentials 1 and 2,
respectively. All of these values are close to the fitted ones at the 
lowest temperature. 

In Fig.~\ref{fig_VT0} we summarize the results for the vacuum potentials 
in the four different color channels which can be formed in 2-body $Q\bar Q$
and $QQ$ systems (recall that in vacuum the entropy-term vanishes and thus 
$F_a=U_a$; also $F_{av}=F_{1}$ from Eq.~(\ref{Fav-def})). The color 
blindness of the string term produces a long-range attraction in all 
channels (which will support colored bound states in vacuum as discussed 
in Sec.~\ref{ssec_vac}). 
The potentials emerging from the model-fit at finite $T$ are collected in 
Fig.~\ref{fig_VT}. One clearly
recognizes the ``melting" of the string term with increasing $T$. The singlet 
(meson) and antitriplet (diquark) potentials remain attractive at all
distances. For the octet and sextet potential some residual attraction
from the string term persist at lower temperatures (especially in $U$), 
preserving a shallow dip structure for quark separations of around 0.1-0.2\,fm.
This behavior has also be seen on the lattice~\cite{Kaczmarek:2005ui} and
is obviously incompatible with a Casimir scaling of the string term.

\section{Thermodynamic $T$-Matrix and Observables}
\label{sec_tmat}
%
\subsection{Reduction Scheme and Relativistic Corrections} 
\label{ssec_red-rel}
The above constructed in-medium potentials are now implemented into a 
thermodynamic $T$-matrix approach. The latter follows from a 
3-dimensional (3D) reduction of the Bethe-Salpeter Equation in ladder 
approximation~\cite{Blankenbecler:1965gx,Thompson:1970wt,Machleidt:1989tm}.
Heavy-quark systems are particularly suitable for this reduction as their 
energy transfer is parametrically suppressed compared to the momentum 
transfer. Even for heavy-light systems the on-shell condition on the 
heavy quark suppresses the energy transfer relative to the 3-momentum 
transfer. Note that a 4D treatment cannot improve the 
accuracy as long as the input is based on static potentials. However, 
relativistic corrections, as well as different reduction schemes, should and 
will be addressed below. The former are necessary to ensure consistency for
relativistic energies  (and, in fact, establish ``minimal" Poincar{\'e}
invariance of the potential approach~\cite{Vairo:xx} (see also \cite{Brambilla:2003nt})), while the latter 
give an indication of uncertainties inherent in the static approximation.  
The 3D integral equation for the $T$-matrix can be further simplified by 
applying a partial-wave decomposition which leads to the following 1D 
equation, 
\begin{eqnarray}
T_{l,a}(E;q^\prime,q)=\mathcal{V}_{l,a}(q^\prime,q)
+\frac{2}{\pi}\int\limits^\infty_0 dk\,k^2\,\mathcal{V}_{l,a}(q^\prime,k)
\qquad
\nonumber\\
\times G_{12}(E;k)\,T_{l,a}(E;k,q)\,[1-n_F(\omega_1(k))-n_F(\omega_2(k))]
 \ , 
\nonumber\\
\label{Tmat}
\end{eqnarray}
for the $T$-matrix $T_{l,a}$ in a given color channel ($a$) and partial 
wave ($l$); $n_F$ is the Fermi-Dirac distribution, $q=|\vec{q}\,|$, 
$q^\prime=|\vec{q}^{\,\prime}\,|$ and $k=|\vec{k}\,|$ denote the relative 
3-momentum moduli of the initial, final and intermediate 2-particle state, 
respectively, and $\omega_i(k)=(m_i^2+k^2)^{1/2}$ the single-quark energies. 
Equation~(\ref{Tmat}) encompasses both the heavy-light ($1=Q$, $2=q$)
and quarkonium ($1,2=Q$) channels where either particle can be 
a quark or an antiquark.  
The precise form of the two-particle propagator, $G_{12}$, depends on the reduction
scheme~\cite{Yaes:1971vw,Frohlich:1982yn}, for which we will investigate two 
well-established options, namely the 
Blancenbecler-Sugar (BbS)~\cite{Blankenbecler:1965gx} and the 
Thompson (Th)~\cite{Thompson:1970wt} scheme,
\begin{eqnarray}
G_{12}^{\rm Th}(E;q)
&=&\frac{m(q)}{E-\omega^q(q)-\omega^Q(q)-\Sigma_q-\Sigma_Q} \ ,\nonumber
\\
G_{12}^{\rm BbS}(E;q)&=&\frac{2\,m(q)\, (\omega^q(q)+\omega^Q(q))}
{E^2-(\omega^q(q)+\omega^Q(q)+\Sigma_q+\Sigma_Q)^2} \ ,\nonumber
\\
m(q)&=&\frac{m_q\,m_Q}{\omega^q(q)\,\omega^Q(q)} \ . 
\label{G2}
\end{eqnarray}
The main difference between both schemes is that the dependence on the total 
energy, $E$, is quadratic for the BbS propagator but linear for the Thompson 
version. The form of the propagators in Eqs.~(\ref{G2}) further 
implies that both quarks remain good quasiparticles in the medium, i.e., 
their widths $\Gamma_{Q,q}$ are small compared to their mass. We use a 
minimal width of $\Gamma_{q,Q}=20$\,MeV to facilitate numerical stability, 
unless otherwise stated. The incorporation of microscopic quark spectral 
functions will be carried out in an upcoming study.
Once the potential is specified, the $T$-matrix equation (\ref{Tmat}) is 
solved using the algorithm of Haftel and Tabakin~\cite{Haftel-NPA158} 
as in previous works in our context~\cite{Mannarelli:2005pz,Cabrera:2006wh}.

\begin{figure}[t]
\includegraphics[scale=1]{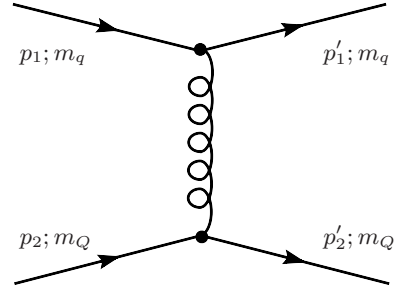}
\caption{One-gluon exchange diagram for quark-quark scattering; in the center of mass system, the
relative 4-momentum in the incoming (outgoing) state is
$q=(p_1-p_2)/2$ ($q'=(p_1'-p_2')/2$).}
\label{fig_oge}
\end{figure}

It remains to specify how we implement the coordinate-space potential as
extracted from lQCD in the previous section. We start by performing the
Fourier transform and partial-wave expansion according to
\begin{eqnarray}
V^{C/S}_{l,a}(q^\prime,q)
&=&\int\frac{d^3r\,dx_{q^\prime q}}{8\pi}P_l(x_{q^\prime q}) 
V^{C/S}_a(r)e^{i(\vec{q}-\vec{q}^{\,\prime})\vec{r}}
\nonumber\\ 
x_{q^\prime q}&=&\frac{\vec{q}^{\,\prime}\cdot\vec{q}}
{\sqrt{|\vec{q}^{\,\prime}\,|\,|\vec{q}\,|}} \ 
\end{eqnarray}
with the usual Legendre Polynomials $P_l$.
Since the string part of the potential ($V^{S}$) is primarily active at
long distances, i.e., at low momenta and thus in the nonrelativistic
regime, no further amendments are applied to it. However, for the Coulomb 
part ($V^{C}$), which dominates at small distances (and thus at relatively
large momentum transfers), several corrections are in order.
To infer relativistic effects, let us back up to the definition 
of the relativistic Coulomb potential given by the perturbative one-gluon 
exchange diagram in Fig.~\ref{fig_oge}.
\begin{figure*}[t]
\includegraphics[scale=1.0]{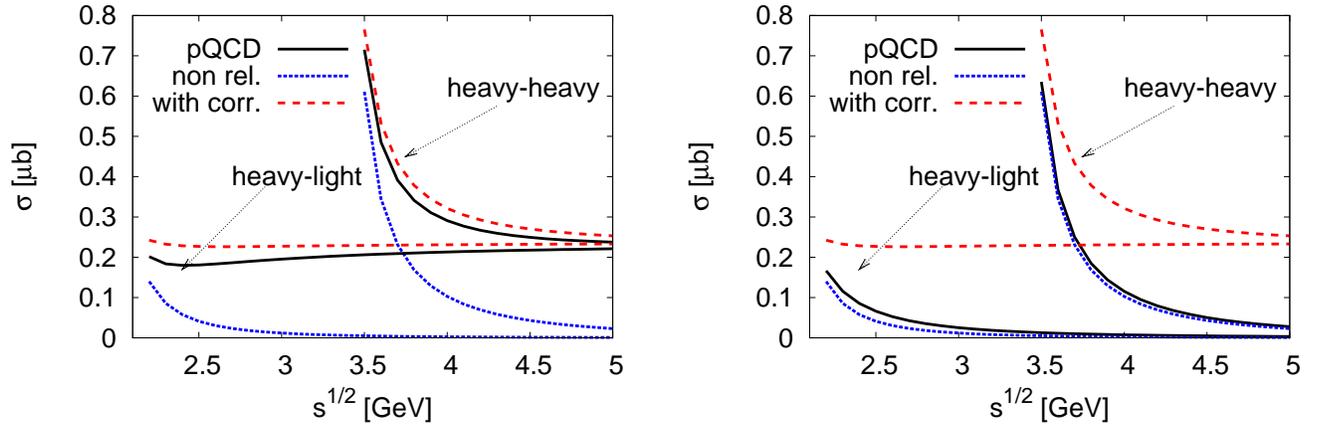}
\caption{(Color online) Left panel: perturbative QCD cross section for 
quark-quark scattering via $t$-channel gluon exchange using a vector 
interaction. The full pQCD result (solid line) is compared to the $T$-matrix
result in Born approximation using (i) a nonrelativistic treatment of the 
spinor structure (doted line) and (ii) the correction factors $R$ and $B$. 
Right panel: the same but for a scalar interaction. 
The heavy- and light-quark masses are set to 1.7~GeV and 0.4~GeV, respectively,
and the strong coupling and Debye mass have been fixed at 0.3 and 0.67~GeV,
respectively.}
\label{fig_pQCD}
\end{figure*}
Suppressing the color structure, the Born amplitude (potential) is given
by
\begin{eqnarray}
V &\sim& \underbrace{\frac{4\pi\alpha_s}{t-m_D^2}}\,
\underbrace{\bar{u}(p_1^\prime)\,\gamma^\mu\,u(p_1)\,\bar{u}(p_2^\prime)\,
\gamma_\mu\,u(p_2)}
\\
&&\,\,\text{Yukawa}\qquad\text{Spinor-contraction}
\nonumber
\end{eqnarray}
where we have included a Debye mass as the leading temperature correction in
the gluon propagator. In addition to the standard Yukawa propagator (which in 
the static approximation amounts to setting $t \to -(\vec q-\vec q\,')^2$ in the center of mass system) 
we have a contribution from the contraction of the spinors with the vertex. 
At the level of the cross section (or amplitude squared) this part gives rise 
to the following factor (with the normalization $\bar{u}\,u=1$), 
\begin{eqnarray}
&&|\bar{u}(p_1^\prime)\,\gamma^\mu\,u(p_1)\,
\bar{u}(p_2^\prime)\,\gamma_\mu\,u(p_2)|^2
\nonumber\\
&&\qquad\qquad=\frac{8(2(s-m_q^2-m_Q^2)^2+2\,s\,t+t^2)}{16\,m^2_q\,m^2_Q} \ .
\label{spinor-vector}
\end{eqnarray}
For large $s=E^2$ the terms proportional $t$ are subleading and can be 
dropped so that we can reformulate this expression as
\begin{eqnarray}
&&|\bar{u}(p_1^\prime)\,\gamma^\mu\,u(p_1)\,\bar{u}(p_2^\prime)\,
\gamma_\mu\,u(p_2)|^2\,
\nonumber\\
&&\qquad\qquad
\simeq\,\frac{8(2(s-m_q^2-m_Q^2)^2)}{16\,m^2_q\,m^2_Q}
\\
&&\qquad\qquad 
=4\frac{\omega^2_q\,\omega^2_Q}{m^2_q\,m^2_Q}
  (1+\frac{q^2}{\omega_q\,\omega_Q})^2
\nonumber
\end{eqnarray}
The factor in parenthesis is precisely the well-known Breit interaction
 in electrodynamics, corresponding to a magnetic current-current
interaction of 2 moving charges~\cite{Brown:2003km,Shuryak:2004tx,Brown:1952},
while the first factor ``corrects" for relativistic kinematics. 
(the extra factor 4 arises from the summation over spins and has to be 
taken out in a spin-independent definition of the potential).
We therefore identify the following factors with which the nonrelativistic 
(off-shell) potential, $V(q,q')$, should be augmented:
\begin{eqnarray}
R(q^\prime,q)&=& m(q)^{-1/2} \ m(q')^{-1/2}   
\\
B(q^\prime,q)&=& b(q)^{1/2} \ b(q')^{1/2}
\\
b(q) &=&  (1+\frac{q^2}{\omega_q(q)\,\omega_Q(q)}) \ .
\end{eqnarray}
Note that $B,R\to 1$ for $q,q'\ll m_{Q,q}$. 
For the string term, for which we assume a scalar interaction, 
the spinor contraction leads to  
\begin{eqnarray}
&&|\bar{u}(q^\prime)\,\,u(q)\,\bar{u}(p^\prime)\,u(p)|^2
\nonumber\\
&&\qquad\qquad=\frac{4(4m_q^2-t)(4m_Q^2-t)}{16\,m^2_q\,m^2_Q} \ .
\label{spinor-scalar}
\end{eqnarray}
Assuming again that we can drop the terms proportional $t$ (relative to 
$m_{Q,q}$), no relativistic correction factor is mandated (this refines the 
procedure adopted in earlier works~\cite{Cabrera:2006wh,Mannarelli:2005pz}).

To check the impact of our corrections we compute the cross sections for 
one-gluon exchange (Fig. \ref{fig_oge}) for quark-quark scattering using 
the Coulomb term in Born approximation including our correction factors,
\begin{eqnarray}
\frac{d\sigma}{d\Omega}=
\frac{1}{64\pi s}\frac{2}{36}\frac{m_q^2\,m_Q^2}{\omega_q^2\,\omega_Q^2}\,
\sum\limits_{l,a}|T_{l,a}|^2 \ \  , \quad T_{l,a}={\mathcal V}^C_{l,a} \ , 
\label{born}
\end{eqnarray}
and compare it to the exact perturbative QCD (pQCD) results in 
the left panel of Fig.~\ref{fig_pQCD}. One finds that the relativistic 
correction factors $B$ and $R$ are essential to establish consistency with
pQCD at high energies; even at low energies, the agreement is no worse
than 20\%, which supports the approximation of neglecting $t$ against
 $s$ and $m_{Q,q}$ in the numerator of Eq.~(\ref{spinor-vector}).
The factors $R$ and $B$ provide a substantial improvement over not 
including them. Without the $R$ factor, one obtains vanishing cross 
sections at high energy (only half of the correct
magnitude without the Breit correction); even close to 
threshold the discrepancy to pQCD is larger than with the corrections. 
Also note that without the relativistic factors the cross section 
goes to zero for $m_q\,\rightarrow\, 0$, which becomes an uncontrolled
feature in applications to heavy-light scattering.  
In the right panel of Fig.~\ref{fig_pQCD} we compare a ''pQCD`` calculation
assuming a scalar vertex structure to the $T$-matrix Born results with and 
without correction factors $B$ and $R$. In this case, it is obviously 
mandated not to include these factors. As to be expected, the 
nonrelativistic approximation leads to the same result irrespective of 
whether one uses a vector or scalar interaction.

Finally, we account for effects of the running coupling constant in the
off-shell extrapolation of the potential. For on-shell kinematics, $q=q'$,
such effects are effectively taken care of by our parametrization of
the potential. For off-shell scattering, we simulate the running coupling
by introducing a factor $F_{\rm run}(q\ne q') <1$ as 
\begin{equation}
F_{run}(q^\prime,q)=\ln\left[\frac{\Delta^2}{\Lambda^2}\right]
/\ln\left[\frac{(q-q^\prime)^2+\Delta^2}{\Lambda^2}\right]
\end{equation}
with $\Delta=1$\,GeV and $\Lambda=0.2$\,GeV.

Putting all corrections together, the final form of the potential
figuring in the $T$-matrix equation (\ref{Tmat}) is
\begin{eqnarray}
\mathcal{V}_{l,a}(q^\prime,q) &=&
R(q^\prime,q)\,B(q^\prime,q)\,F_{run}(q^\prime,q)\,V^{C}_{l,a}(q^\prime,q)
\nonumber\\
&&+V^{S}_{l,a}(q^\prime,q) \ . 
\label{potential}
\end{eqnarray}
In vacuum the unscreened Coulomb potential exhibits a well-known infrared
singularity. We tame this by introducing a small low-momentum cutoff; 
we have checked that varying this cutoff has a very small effect
on the vacuum spectral functions. 
%
\subsection{Quarkonium Correlators and HQ Diffusion}
The key quantity for computing observables is the on-shell $T$-Matrix,
$T_{l,a}(E;q,q)$, where $E=\omega_1(q)+\omega_2(q)$ for both in- and outgoing
channels. Following Ref.~\cite{Cabrera:2006wh}, the continuation below the 
2-particle threshold, $E_{\rm thr}=m_1+m_2$, is carried out for vanishing 
3-momentum, $T_{l,a}(E;0,0)$. 
The mesonic spectral function in a given quantum-number channel $\alpha$
is defined as 
\begin{eqnarray}
\sigma_\alpha(E)=-\frac{1}{\pi}\,{\rm Im}\,\mathcal{G}_\alpha(E)\,,
\label{sf}
\end{eqnarray}
where $\mathcal{G}$ denotes the correlation function, for which we will focus
on the case of a heavy quark and antiquark  ($Q\bar Q$) in the color-singlet
channel ($a$=1) in a QGP of vanishing net baryon charge ($\mu_q$).
It can be written as 
\begin{equation}
\mathcal{G}_\alpha(E)=\mathcal{G}^0_{\alpha}(E)+\Delta \mathcal{G}_a(E)
\label{corr}
\end{equation}
where 
\begin{eqnarray}
\mathcal{G}^0_{\alpha}(E)&=&i N_f N_c \int\frac{d^3k}{(2\pi)^3}\,G_{12}(E;k)\,
\left[1-2n_F(\omega_Q(k)\right]
\nonumber\\
&&\qquad\times {\rm Tr}\left[\Gamma_\alpha\,\Lambda_+(\vec{k}\,)\,
\Gamma_\alpha\, \Lambda_-(-\vec{k}\,)\,\right]
\label{corr-0}
\end{eqnarray}
denotes the noninteracting contribution with
the particle/antiparticle projectors
\begin{equation}
\Lambda_\pm(\vec{k}\,)=\frac{\omega_Q(k)\gamma^0-\vec{k}\cdot
\vec{\gamma}\pm m_Q}{2\,m_Q} \ . 
\end{equation}
The Dirac matrices $\Gamma_\alpha \in 
\{1,\gamma_\mu, \gamma_5, \gamma_\mu\gamma_5\}$ characterize the scalar,
vector, pseudoscalar and pseudovector channels, respectively (corresponding 
to $\chi_{c0}$, $J/\psi$, $\eta_c$ and $\chi_{c1}$ in the charmonium sector).
In the following we will neglect effects due to spin-orbit and spin-spin
(hyperfine) interactions which is justified in the HQ limit. It implies
degeneracy of the $S$-wave ($l$=0) states $J/\psi$ and $\eta_c$, as well
as of the  $P$-wave ($l$=1) $\chi_c$ states (in the vacuum spectrum, this 
is realized within $\sim$0.1\,GeV). Thus, we will compute only the vector 
($\Gamma_\alpha=\gamma_\mu$) and scalar ($\Gamma_\alpha=1$) channels.
The interacting contribution to the correlator in Eq.~(\ref{corr}) is given 
in terms of the off-shell $T$-matrix as 
\begin{eqnarray}
\Delta \mathcal{G}_\alpha(E)
&=&\frac{N_f N_c}{\pi^3}\int dk\,k^2 \,G_{12}(E;k)\,[1-2\,n_F(\omega_Q(k))]
\nonumber\\
&&\hspace{-1.0cm}\times 
\int dk^\prime\,{k^\prime}^2 \,G_{12}(E;k^\prime)\,
[1-2\,n_F(\omega_Q(k^\prime))]\,
\nonumber\\
&&\hspace{-1.0cm}\times 
\left[a_0(k,k^\prime)\,T_{0,a}(E;k,k^\prime)+a_1(k,k^\prime)\,
T_{1,a}(E;k,k^\prime)\right]\,.
\nonumber \\
\label{corr-int}
\end{eqnarray}
The coefficients $a_l$ result from the traces over the spinor structure. 
In line with the above approximation of neglecting spin-induced interactions,
we use a $1/m_Q$ expansion for these coefficients (which also leads to the
degeneracy of  pseudoscalar-vector and scalar-axialvector). One 
has~\cite{Cabrera:2006wh}
\begin{eqnarray}
a_0=2\qquad a_1=0
\end{eqnarray}
for $S$ waves and 
\begin{eqnarray}
a_0=0\qquad a_1=-2\frac{k\,k^\prime}{m_Q^2} 
\end{eqnarray}
for $P$ waves. 

To test or quarkonium spectral functions against lQCD correlators, computed
with good accuracy in euclidean space-time, we need to calculate the 
euclidean-time correlator defined as
\begin{eqnarray}
G_\alpha(\tau,T)&=&\int dE\,\sigma_\alpha(E,T)\,
\mathcal{K}(\tau,E,T) \ ,
\\
\mathcal{K}(\tau,E,T)&=&\frac{\cosh\left[E\,\left(\tau-\beta/2\right)
\right]}{\sinh\left[E\beta/2\right]} \ .
\nonumber
\label{eucl-corr}
\end{eqnarray}
The use of a constant width in the two-particle propagator,
Eqs.~(\ref{G2}), implies an non-vanishing value 
for $\sigma_\alpha(E\to0)$. This induces an artificial singularity in the 
euclidean correlator since the temperature kernel, $\mathcal{K}$, diverges
in the zero-energy limit. This is an artifact of the quasiparticle
approximation that can in principle be cured by employing a microscopic
calculation of the single-particle selfenergies leading to the proper
limit, $\sigma_\alpha(E\to0)\to E$, for the retarded meson spectral function. 
We defer this study to future work and evade the singularity problem 
by imposing a cutoff below which we set the imaginary part of the propagator 
to zero, $E_{\rm cut}=2(8)$~GeV for charmonia (bottomonia); there is very
little sensitivity to our correlator results when decreasing $E_{\rm cut}$
by up to a factor of 2. 

An additional subtlety in the comparison of model spectral functions to
lQCD euclidean correlators~\cite{Umeda:2007hy,Aarts:2005hg} is the presence
of so-called zero-mode (ZM) contributions. These may be
pictured as changing the time direction of a HQ line and thus represent 
HQ scattering processes including Landau damping (rather than $Q\bar Q$ 
propagation). 
It turns out that the pseudoscalar quarkonium channel does not pick up 
the ZM contribution. To avoid extra uncertainties due to the latter,
we therefore restrict our comparisons to euclidean lQCD correlators
to this channel (recall that within our approximations the $S$-wave 
pseudoscalar ($\eta_c$) and vector ($J/\psi$) channels are degenerate).  
A common way to display the euclidean correlator at a given temperature
uses a normalization to a so-called reconstructed correlator, which
uses a baseline spectral function (e.g. the vacuum one) with
the Kernel $\mathcal{K}$ at the same temperature as in numerator,  
\begin{eqnarray}
R_\alpha(\tau;T)=\frac{\int dE\,\sigma_\alpha(E,T)\,
\mathcal{K}(\tau,E,T)}{\int dE\,\sigma_\alpha(E,T_{\rm rec})\,
\mathcal{K}(\tau,E,T)} \ .
\label{RG}
\end{eqnarray}
The idea underlying this ratio is to exhibit the temperature effects on the 
in-medium spectral function, $\sigma_\alpha(E,T)$ relative to a baseline 
spectral function, $\sigma_\alpha(E,T_{\rm rec})$, and thus to reduce the 
effects of systematic uncertainties (e.g. discretization effects in lQCD
which distort the high-energy part of the spectral functions).
As pointed out in Ref.~\cite{Cabrera:2006wh} the spectral function used
in the reconstructed correlator can have significant impact on the
normalization and shape of $R_\alpha(\tau;T)$. Here, we always use our 
calculated vacuum spectral function as baseline, i.e., $T_{\rm rec}$=0.

Let us finally elaborate on the diffusion properties of a single heavy 
(anti-) quark which we evaluate in terms of our heavy-light quark $T$-matrix.
Following Ref.~\cite{Svetitsky:1987gq}, one may approximate the Boltzmann
equation for the HQ distribution function in the QGP by a Fokker-Planck
equation and extract the pertinent thermal relaxation rate (inverse
relaxation time) as 
\begin{equation}
\gamma_c=1/\tau_{Q} \equiv \lim\limits_{p\to0} A(\vec{p}\,)
\end{equation}
with the friction coefficient 
\begin{eqnarray}
&A(\vec{p}\,)&=\frac{1}{16\,(2\pi)^9\,\omega_Q(p)}
\int\frac{d^3q}{\omega_q(q)} \ n_F(\omega_q(q))
\int\frac{d^3q^\prime}{\omega_q(q^\prime)} 
\nonumber\\
&&\times\int\frac{d^3p^\prime}{\omega_Q(p^\prime)}
\frac{(2\pi)^4}{d_c}\, \sum|M|^2\,\delta^{(4)}(q+p-q^\prime-p^\prime) 
\nonumber\\
&&\qquad\times \left(1-\frac{\vec{p}\cdot\vec{p}^{\,\prime}}{\vec{p}^{\,2}}\right) \ .
\label{Ap}
\end{eqnarray}
The invariant amplitude squared, which is summed over color, angular-momentum,
spin and light-flavor degrees of freedom (and averaged over the $d_c=6$ 
initial spin-color states of a heavy quark), is calculated in terms of 
$S$- and $P$-wave on-shell $T$-matrices as 
\begin{eqnarray} 
\sum|M|^2=\frac{64\pi}{s^2}(s-m_q^2+m_Q^2)^2(s-m_Q^2+m_q^2)^2
\nonumber\\
\times\,N_f\sum\limits_a d_a(|\tilde{T}_{a,0}(s)|^2
+3|\tilde{T}_{a,1}(s)\,\cos(\theta_{cm})|^2)
\end{eqnarray}
where\footnote{Due to the slightly different definition of the relativistic 
factors in our $T$-matrix compared to Ref.~\cite{vanHees:2007me} the 
connection to the cross section is modified~\cite{Herrmann:1992zz}.} 
\begin{equation}
\tilde{T}_{a,i}(s)\equiv
m(p_{\rm cm})^{1/2} \ T_{a,i}(E,p_{\rm cm},p_{\rm cm}) \ m(p_{\rm cm})^{1/2}
\end{equation}
with center-of-mass (cm) energy and momentum
\begin{eqnarray}
E&=&\sqrt{s}=\omega_q(p_{\rm cm})+\omega_Q(p_{\rm cm})
\nonumber\\
p_{\rm cm}&=&\frac{1}{2E}\,\sqrt{m_Q^4 + (m_q^2-s)^2 - 2 m_Q^2 (m_q^2+s)} \ .
\nonumber
\end{eqnarray}
The color degeneracy factors are given by 
\begin{equation}
d_0=1,\quad d_{\bar{3}}=3,\quad d_6=6,\quad d_8=8 \ .  
\end{equation}
In Eq.~(\ref{Ap}), the distribution functions, $n_F$, include up ($u$), 
down ($d$) and strange ($s$) quarks in the thermal heat bath with 
incoming (outgoing) 3-momentum, $\vec q$ ($\vec q^{\,\prime}$). 
The in- and outgoing HQ 3-momenta are $\vec p$ and $\vec{p}^{\,\prime}$.
As an extension to previous work~\cite{vanHees:2004gq,vanHees:2007me} we 
here distinguish explicitly between light- and strange-quark contributions 
(instead of using light quarks with an effective degeneracy of $N_f=2.5$).
In accordance with HQ spin symmetry (as adopted in the quarkonium sector)
we assume degeneracy of $S$-waves (e.g., pseudoscalar $D$ and vector $D^*$ 
mesons) and of $P$-waves (e.g., scalar $D_0$ and axialvector $D_1$ mesons). 
In our numerical calculations below we also evaluate the contributions 
from HQ scattering off gluons. In this case, a potential picture is less 
straightforward. Therefore, as in previous 
work~\cite{vanHees:2004gq,vanHees:2007me}, we account for elastic 
HQ-gluon scattering via the leading order perturbative diagrams 
(including a Debye screening mass) with a rather large coupling constant 
of $\alpha_s=0.4$. 

\section{Spectral Functions, Correlators and Relaxation Times}
\label{sec_sf}
In this Section we first fix the remaining free parameters, i.e., 
the bare heavy- as well as light- and strange-quark masses and
check the resulting spectral functions against vacuum spectra of
hidden and open heavy-flavor mesons (Sec.~\ref{ssec_vac}). 
We then discuss our numerical results for quarkonium spectral functions
and pertinent euclidean correlators with emphasis on the uncertainties
originating from the choice of potential and reduction scheme 
(Sec.~\ref{ssec_onia}). Finally, we apply our formalism to evaluate 
HQ thermalization times and diffusion coefficients 
(Sec.~\ref{ssec_HQ-diff}).   

\subsection{Vacuum Spectroscopy and Quark Masses}
\label{ssec_vac}
%
\begin{table}[!b]
\begin{tabular}{|l|l|l|l|}
\hline
&  & BbS scheme & Th scheme \\
\hline
Potential 1 & $m_c^0$ & 1.355 GeV& 1.264 GeV\\
            & $m_b^0$ & 4.712 GeV& 4.662 GeV\\
\hline
Potential 2 & $m_c^0$ & 1.402 GeV& 1.293 GeV\\
            & $m_b^0$ & 4.768 GeV& 4.700 GeV\\
\hline
\end{tabular}\\
\begin{center}
$m_q=0.4$ GeV,  $m_s=0.55$ GeV
\end{center}
\caption{Dependence of the bare HQ masses on the 3-D reduction scheme
and underlying lattice potential when adjusting the ground-state quarkonium
mass to the average experimental value. Also quoted are the effective
light- and strange-quark masses which have been adjusted to
the average heavy-light meson ground-state mass.}
\label{tab_mass}
\end{table}
\begin{figure*}[!t]
\includegraphics[scale=0.62]{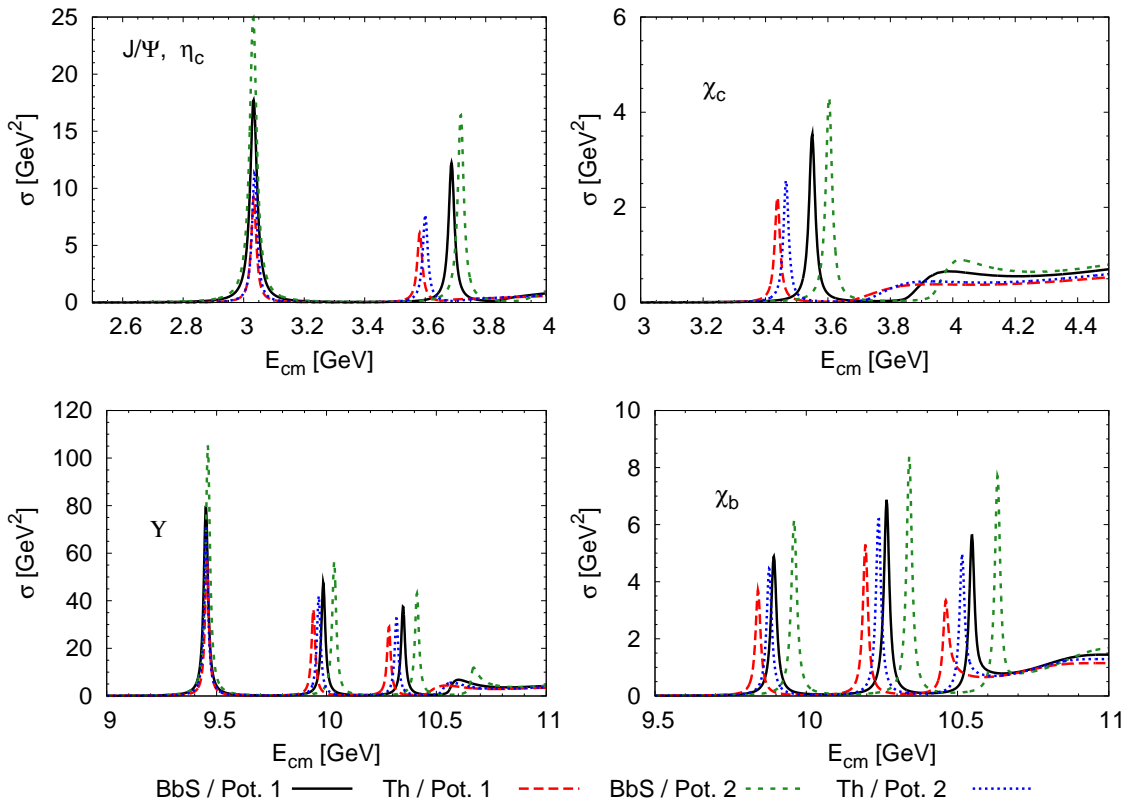}
\caption{(Color online) Quarkonium spectroscopy in vacuum for
$S$- and $P$-wave charmonia (upper panels) and $S$- and $P$-wave
bottomonia (lower panels).}
\label{fig_onium-vac}
\end{figure*}

Let us first focus on the vacuum spectra of charmonia and bottomonia to
determine the bare masses of charm ($c$) and bottom ($b$), $m_{c,b}^{0}$,
which figure into the expression for the effective mass, Eq.~(\ref{m-eff}).
We do this by requiring the $S$-wave charmonium (bottomonium) ground state 
to occur at the average mass of $\eta_c$ and $J/\psi$ at $\sim$3.04~GeV,
and at the $\Upsilon(1S)$ mass of $\sim$9.46~GeV (we neglect hyperfine
splittings), see Fig.~\ref{fig_onium-vac}.
Since the entropy term in the HQ free energy vanishes in the vacuum
there is no difference between the free and internal energy. The resulting 
bare-quark masses are compiled in Tab.~\ref{tab_mass}
for the two different potentials and reduction schemes. They generally fall
into the range expected from the bare masses quoted by the particle data
group~\cite{pdg08} and are also consistent with previous $T$-matrix
calculations~\cite{Cabrera:2006wh}.
The spread (in particular the relative
one) is somewhat larger in the charm sector (ca.~140\,MeV) than in the
bottom sector (ca.~100\,MeV), in line with the expectation that the 
3-D reduction becomes more reliable with increasing mass. 
The mass gap between the ground and first excited charmonium state varies 
rather little between the two potentials within a given reduction scheme,
$\delta m_\psi$=0.65-0.68\,GeV (BbS) and 0.54-0.56\,GeV (Th). Compared
to the experimental value of $m_{\psi'}-m_{J/\psi}=0.59$\,GeV, the
Thompson scheme seems to be doing slightly better (the bare masses 
in the BbS scheme tend to be slightly high).  
The situation is opposite for the pseudoscalar mass splitting between
$\eta_c$ and $\eta_c(2S)$, where the BbS scheme does slightly better
(however, the effects of the hyperfine splitting are expected to be
larger in the pseudoscalar than in the vector case).
The Th scheme appears to perform somewhat better again for the $\chi_c$ 
states, for which the BbS scheme overpredicts the spin-averaged mass
by up to 0.1\,GeV. From these observations we deduce an overall uncertainty
of our $T$-matrix approach of 50-100\,MeV in the charmonium sector,
comparable to corrections one expects from hyperfine splittings. 

\begin{figure*}[!t]
\includegraphics[scale=0.62]{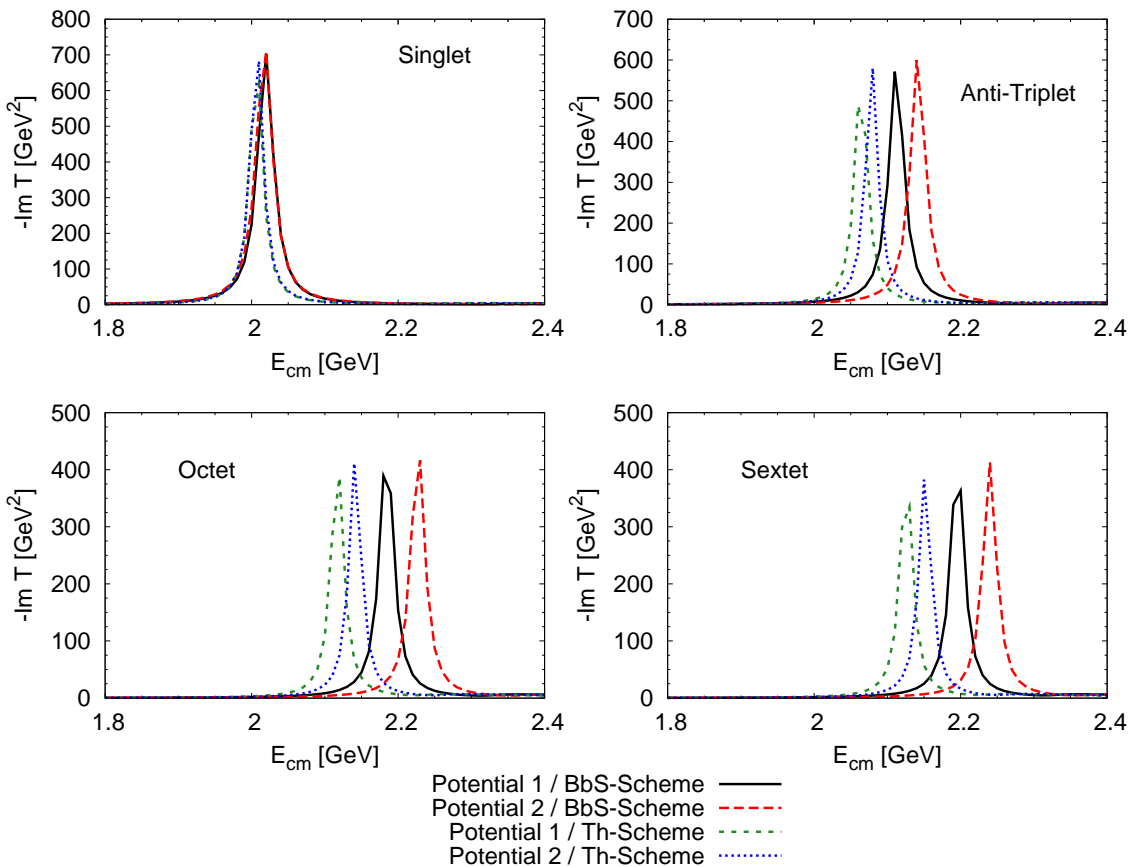}
\caption{(Color online) Imaginary part of the charm-light $T$-Matrix
in the vacuum using $m_q=0.4$\,GeV. We show the color singlet $Q\bar q$
(upper left plot), anti-triplet $Qq$ (upper right plot), octet $Q\bar q$
(lower left plot) and sextet $Qq$ (lower right plot) channels.}
\label{fig_HLvac}
\end{figure*}
In the bottomonium sector (lower panels in Fig.~\ref{fig_onium-vac}),
the mass gaps between the ground state $\Upsilon$ and the first exited 
state, as well as between the first and second exited state, are 
reproduced within 30\,MeV (BbS) and 70\,MeV (Th). The differences in 
the potentials have again only minor impact. A similar trend is found
for the spin-averaged masses of the $\chi_b$ states:
using the BbS scheme our results tend to be higher in mass (especially 
for potential 2) compared to the experimental
values for $m_{\chi_{b0}}$ and $m_{\chi_{b2}}$, while for the Th scheme 
we typically obtain results 30\,MeV below experiment. In addition, for 
both reduction schems and potentials, we obtain a $\chi_b(3S)$ state 
right at the continuum threshold. Since there is no experimental evidence 
for this state, this could again be indicative for some over-binding. 
Recall, however, that we do not 
account for residual $B$-$\bar B$ interactions in our single-channel 
treatment, which could have a significant impact on the spectral function 
especially close to threshold. As to be expected, 
the difference in the Coulomb term of the two vacuum potentials 
(different $\alpha_s$ but equal string tension) induces larger
deviations for the more deeply bound bottomonia, while the sensitivity
to the reduction scheme (static approximation) is reduced. 
Overall, the accuracy of the predictions of our $T$-matrix approach 
is at the 10\% level of the $1S$-$2S$ mass splittings. This is of
the same order (or even below) the observed hyperfine splittings. This
seems reasonable given that we have neglected both spin-spin and spin-orbit 
interactions at the quark level, as well as residual mesonic interactions
in $D\bar D$ and $B\bar B$ channels.   

Finally, the values of light- and strange-quark mass have to be
fixed. Since the physics of their effective vacuum masses is rather
different than in the HQ sector (spontaneous chiral symmetry breaking
vs. string breaking), we directly adjust the constituent masses.   
With $m_q=0.4$\,GeV we obtain a $S$-wave $D$-meson mass of
2.01(2.02)\,GeV in the Th (BbS) scheme which coincides with the
experimental value for the $D^*$ meson (but is larger than the average 
$D$-$D^*$ mass by ca.~60-70\,MeV), see Fig.~\ref{fig_HLvac}. It turns 
out that both smaller and larger $m_q$ lead to a larger $D$-meson mass: 
in the former case the increase in kinetic energy dominates, while in 
the latter case the increase in mass is more important. 
The result for the $D$-meson mass is roughly consistent with
the string-breaking scale in the HQ potential, in the sense that the 
$D\bar D$ threshold (= twice the $D$-meson mass) approximately
coincides with twice the the separation energy of the $Q\bar Q$ pair
plus their bare masses, 
\begin{equation}
2m_D \simeq V(r_{\rm SB}) + 2m_c^0 = 2 m_c \ .
\end{equation} 
In this interpretation, the binding energy of the heavy-light system 
should coincide with the constituent light-quark mass. This is roughly
satisfied in the charm sector ($m_D$ is $\sim$3-10\% larger than $m_c$) 
while the agreement improves in the bottom sector. 
Interesting effects are found in the non-singlet color channels (which
will figure into our calculations of HQ transport in Sec.~\ref{ssec_HQ-diff}
below), cf.~Fig.~\ref{fig_HLvac}. In the color-antitriplet diquark channel,
where the Coulomb term brings in half of the attraction as in the 
mesonic (color-singlet) channel, a bound state is observed at about
$m_{Qq}\simeq2.1\pm0.05$\,GeV, corresponding to a binding energy of 
ca.~0.15\,GeV. To construct a charmed baryon, one may imagine to add 
another light quark with an estimated binding of $\sim$0.25\,GeV, in 
analogy to the $D$-meson. The resulting baryon mass would amount to 
$\sim$2.25\,GeV, not far from the empirical $\Lambda_c$ mass of 2.29\,GeV.
The color-Coulomb is repulsive in the sextet and octet channels, implying
that the states at around $\sim$2.2\,GeV are entirely due to the confining
force. It is tempting to speculate that the binding of an octet and 
anti-octet (or sextet and anti-sextet), with a binding energy comparable 
to the ground-state charmonium, $\sim$0.6\,GeV, could be a relevant 
configuration underlying the recently discovered $X$, $Y$ and $Z$ states 
in the 3.8-4.5\,GeV mass region. The small widths of these states would be 
naturally explained due to their predominantly color non-singlet building 
blocks, see also Refs.~\cite{Maiani:2004vq,Maiani:2005pe,Ebert:2008kb}.  
If this picture is correct, one predicts further regimes of rich spectroscopy 
for narrow ``exotic" 4-quark states around masses of 6\,GeV (2$c$2$\bar c$), 
10\,GeV ($b\bar bq\bar q$, $2b2\bar q$, $2\bar b2q$) and 20\,GeV 
(2$b$2$\bar b$).   

The empirical heavy-strange mesons, $D_s$ and $D_s^*$, are ca.~100\,MeV 
heavier than the non-strange states ($D$ and $D^*$). We can reproduce
this splitting by choosing a constituent strange-quark mass of
$m_s=0.55$\,GeV, consistent with typical values in constituent quark
models. Other properties of the $cs$ states are quite similar to our 
findings for the $cq$ states and will not be reiterated here. This
also applies to the open-bottom $bq$ and $bs$ states.

\subsection{Quarkonium Spectral and Correlation Functions in the 
Quark-Gluon-Plasma}
\label{ssec_onia}
With all parameters fixed and in-medium potentials determined we now
proceed to compute the spectral functions of heavy quarkonia in the
QGP. These can be tested by comparing the pertinent euclidean correlator
ratios, Eq.~(\ref{RG}), to computations of this quantity on the lattice. 
Recent results by Jakov\'{a}c et al.~\cite{Jakovac:2006sf} in quenched QCD 
and by Aarts et al.~\cite{Aarts:2007pk} for $N_f=2$ show small variations 
of about 10\% of the correlator ratios for charmonia up to temperatures 
of about 2\,$T_c$, and even less for bottomonia. Such a behavior could 
be semi-quantitatively reproduced in several potential model 
approaches~\cite{Wong:2006bx,Alberico:2006vw,Cabrera:2006wh,Mocsy:2007yj}.
However, no systematic assessment of relativistic corrections has been 
performed in these works. 

We limit our in-medium investigations to the temperature regime 
$T\ge1.2\,T_c$; closer to $T_c$, the infnite-distance limit of the internal
energy, $U_\infty(T)$,  exhibits a rapid increase which is presumably 
associated with the onset of phase-transition dynamics. We do not expect 
pertinent effects to be properly accounted for in our current 
single-channel ($Q\bar Q$) implementation of the $T$-matrix. E.g., close 
to a second-order phase transition, long-range many-body correlations 
become important, as well as new degrees of freedom such as $D\bar D$ 
channels, which are not included here.

In the following, we divide the presentation into the charmonium 
(Sec.~\ref{sssec_charmonium}) and bottomonium 
(Sec.~\ref{sssec_botttomonium}) sectors, followed by a combined
evaluation (Sec.~\ref{sssec_disc}).

\subsubsection{Charmonium}
\label{sssec_charmonium}
%
\begin{figure*}[t]
\includegraphics[scale=0.62]{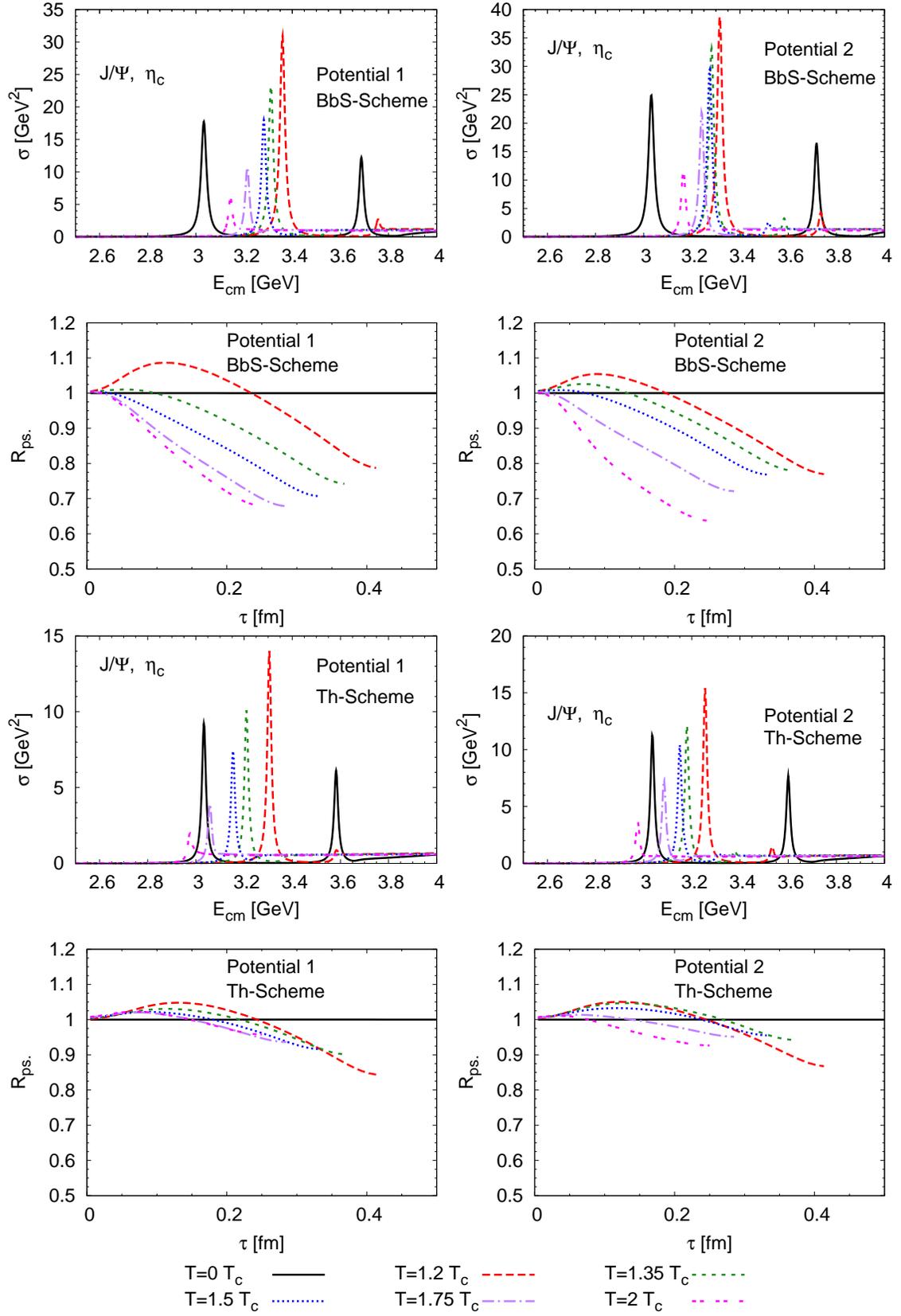}
\caption{(Color online) Charmonium spectral functions and euclidean 
correlators in the pseudoscalar ($S$-wave) channel at various temperatures 
using the internal energy ($U$) as potential. Results for two reduction 
schemes and two different potentials are compared.}
\label{fig_PS-U}
\end{figure*}
\begin{figure*}[t]
\includegraphics[scale=0.62]{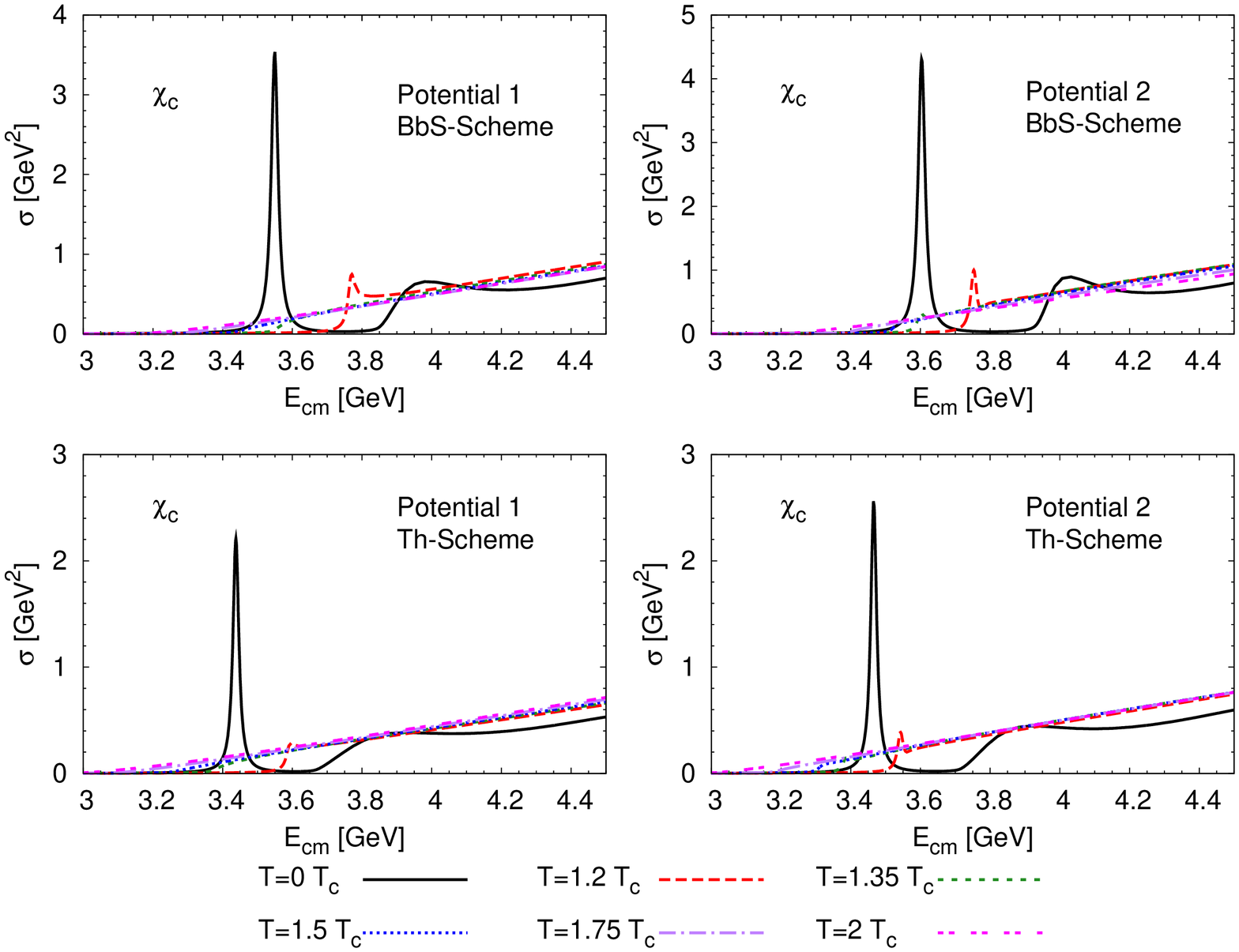}
\caption{(Color online) Charmonium spectral functions in the scalar 
($P$-wave) channel at various temperatures
using the internal energy ($U$) as potential. Results for two reduction
schemes and two different potentials are compared.}
\label{fig_S-U}
\end{figure*}
\begin{figure*}[t]
\includegraphics[scale=0.62]{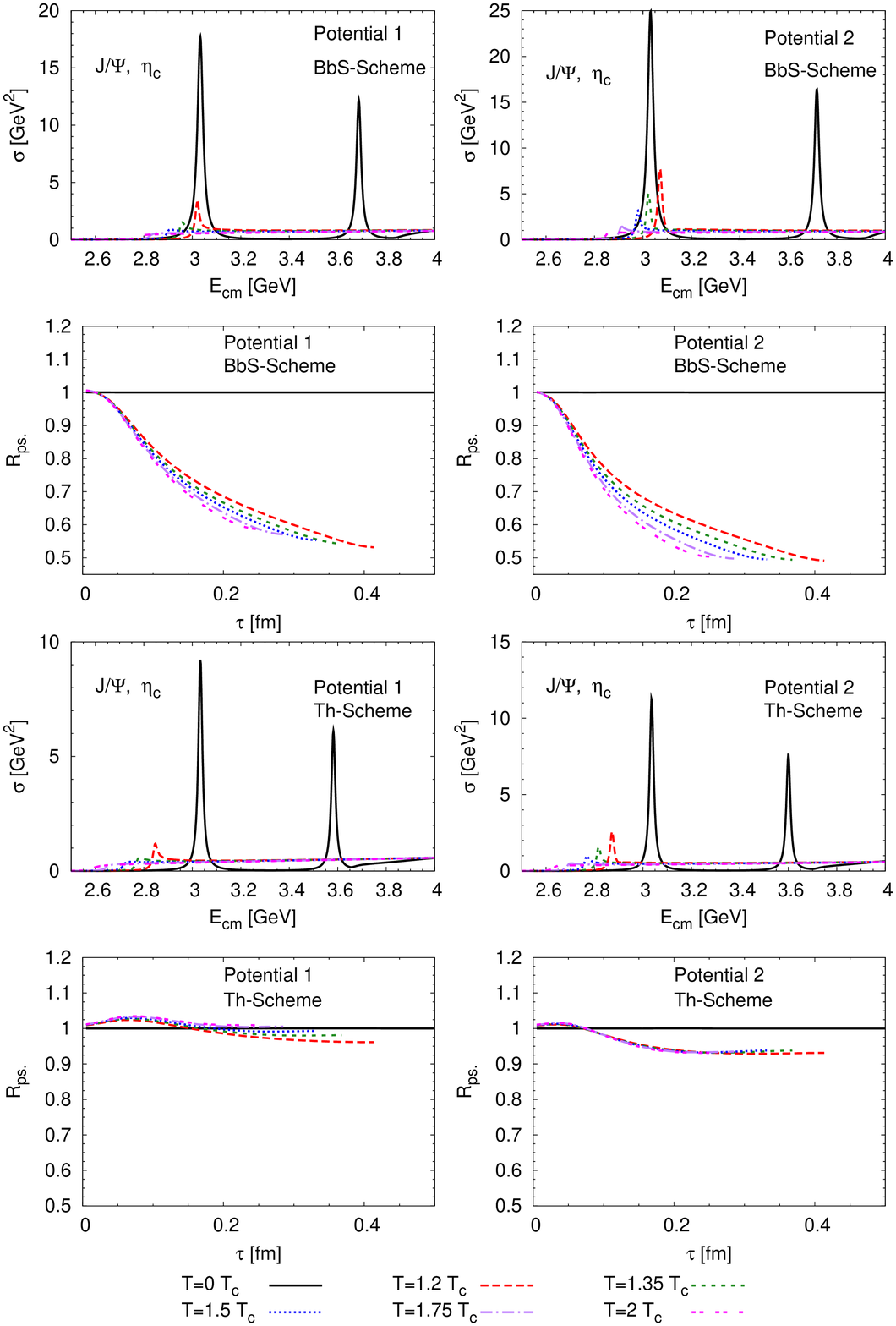}
\caption{(Color online) Same as Fig.~\ref{fig_PS-U} but
using the free energy ($F$) as potential.}
\label{fig_PS-F}
\end{figure*}
\begin{figure*}[t]
\includegraphics[scale=0.62]{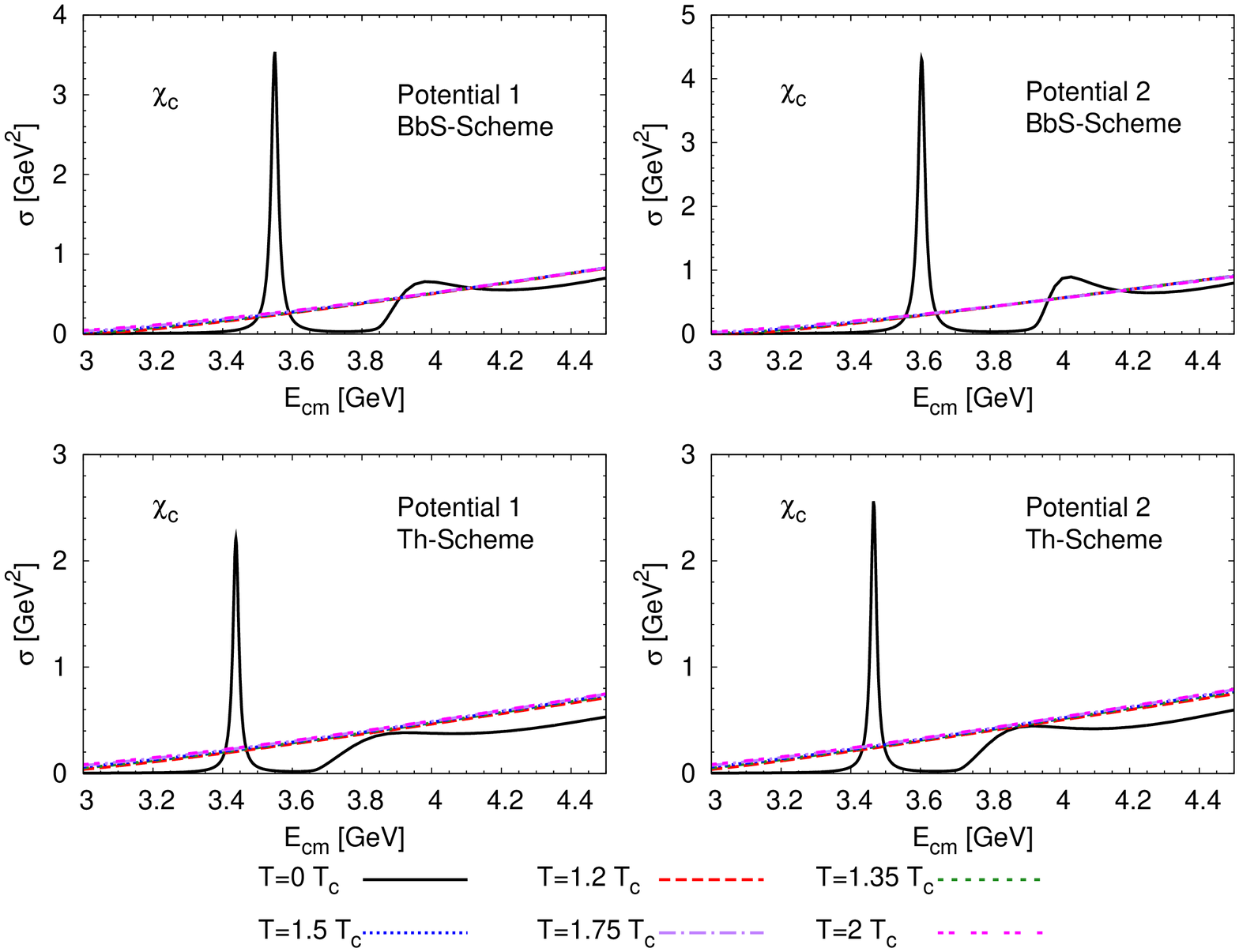}
\caption{(Color online) Same as Fig.~\ref{fig_S-U} but
using the free energy ($F$) as potential.}
\label{fig_S-F}
\end{figure*}
\begin{figure*}[t]
\includegraphics[scale=0.62]{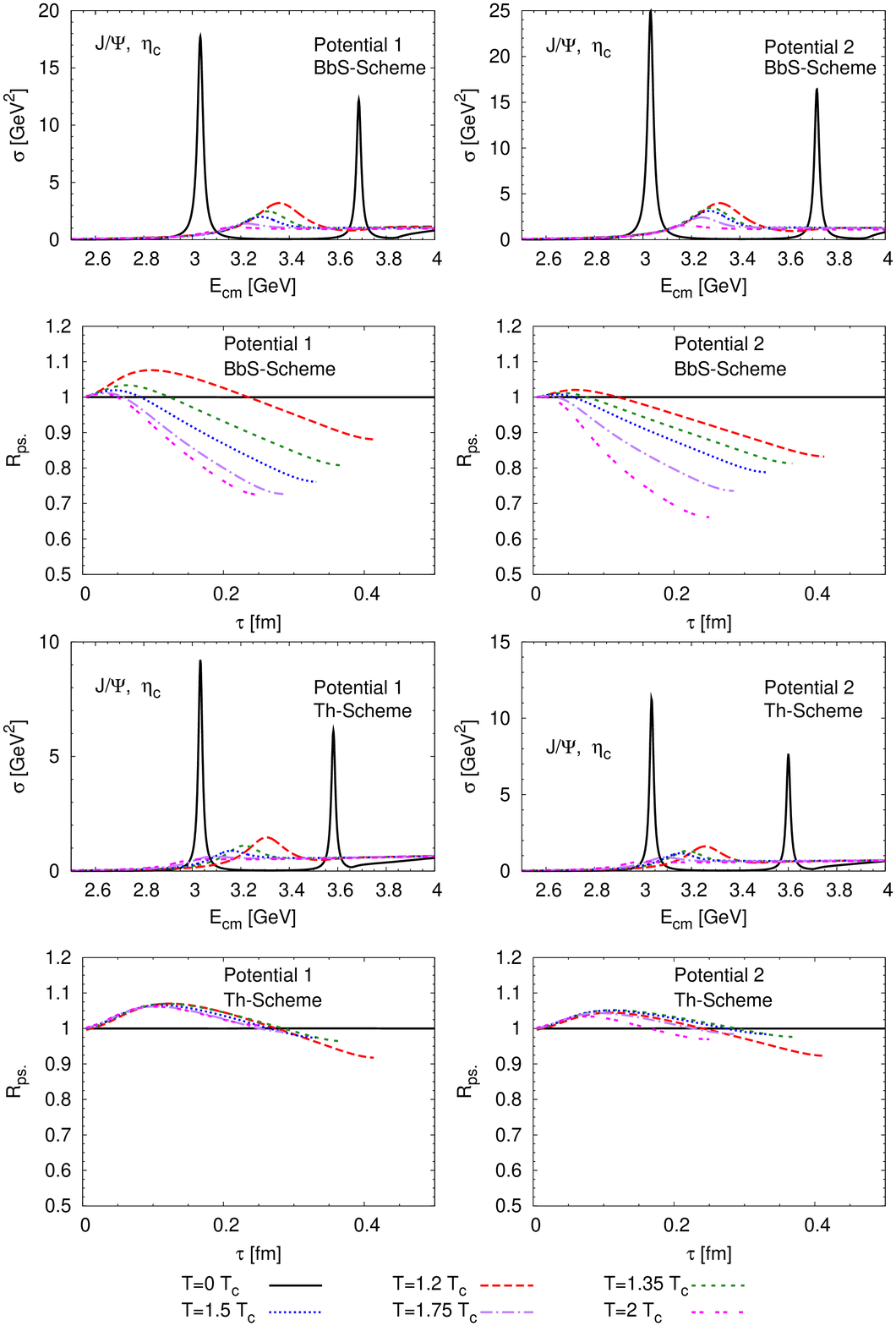}
\caption{(Color online) Same as Fig.~\ref{fig_PS-U} but
employing a single-quark width of 100\,MeV.}
\label{fig_PS-U-width}
\end{figure*}
We begin our in-medium analysis for the $S$- and $P$-wave channels in the
charmonium sector using a small ``numerical width" of 20\,MeV for the $c$
and $\bar c$ quarks (recall the degeneracy of pseudoscalar-vector as well as
of scalar-axialvector states).
Contrary to the vacuum, we now distinguish 2 scenarios depending 
on whether the free ($F$) or internal ($U$) energy is identified as the
static finite-temperature potential. The results are compiled in 
Figs.~\ref{fig_PS-U}+\ref{fig_S-U} for $U$ and in 
Figs.~\ref{fig_PS-F}+\ref{fig_S-F} for $F$ as potential. 

Let us first focus on the former case, $V(r;T)=U(r;T)-U_\infty(T)$. At the
level of the in-medium spectral functions both lQCD inputs and reduction 
schemes share a number of generic trends, all of which were already present
in the $T$-matrix calculations of Ref.~\cite{Cabrera:2006wh}. The $S$-wave 
ground state 
($\eta_c$, $J/\psi$) survives as a bound state up to temperatures of about 
2-2.5\,$T_c$ around which it merges into the $c\bar c$ continuum. But even 
at temperatures as low as 1.2\,$T_c$ the medium effects in the potential 
induce a reduction of the binding energy, $E_B=2m_c-m_\psi$ by about a 
factor of $\sim$2, to $E_B\simeq$~0.3-0.4\,GeV compared to 0.6-0.8\,GeV in 
the vacuum (for Th and BbS, respectively). The effective quark mass at 
this temperature is approximately the same as in vacuum, 
causing a net increase in the mass of the $S$-wave ground state to
$m_\psi(1.2T_c)\simeq$~3.3-3.4\,GeV. For higher temperatures, the binding
further decreases but this effect is overcompensated by a reduction in the
effective charm-quark mass (i.e., in $U_\infty(T)/2$), so that the mass
of the state actually decreases. Along with the decrease in binding goes
a reduction in the strength of the state (= peak height of the spectral 
function at fixed width).  
The rather subtle differences in the spectral functions become more apparent 
in the euclidean correlator ratios, especially between the two reduction 
schemes (within a given reduction scheme, the two different potentials lead 
to small variations also for the correlator ratios). For the BbS scheme, 
the ratios deviate by up to 30-40\% from one for temperatures of 
1.2-2\,$T_c$. This is 
too large compared to the 10-15\% reduction that has been found 
in lQCD computations~\cite{Datta:2003ww,Jakovac:2006sf,Aarts:2007pk}.
However, employing the Thompson scheme, the correlator ratios 
are within 15\% from one, which is better in line with lQCD. 
The technical reason for the difference in the correlator ratios between 
BbS and Thompson scheme can be traced back to the larger binding that the 
BbS scheme generates already in the vacuum. This requires relatively
large bare charm-quark masses (recall Tab.~\ref{tab_mass}) which in the 
medium ultimately lead to too large a ground-state mass (or continuum
threshold) when the latter approaches its dissolution (note that in the BbS
scheme the $J/\psi$ (or $\eta_c$) mass at 2\,$T_c$ is still significantly
above its vacuum mass, while in the Th scheme it has dropped below the
vacuum value). 
For the ground-state $P$-wave state ($\chi_c$) we also find that, right 
above $T_c$, it is heavier than in vacuum due to the decrease in binding. 
However, due to its relatively the small binding energy (in vacuum 
$E_B\simeq$~0.22-0.25\,GeV and 0.3-0.35\,GeV in the Th- and BbS scheme,
respectively) it dissolves just above $\sim$1.2\,$T_c$ where it merges
into the $c\bar c$ continuum.  

Next we discuss the in-medium charmonium results when using $F$ as potential, 
$V(r;T)=F(r;T)-F_\infty(T)$, summarized in Figs. \ref{fig_PS-F} and 
\ref{fig_S-F}. Compared to the use of $U$, the in-medium binding is 
appreciably reduced (recall Fig.~\ref{fig_VT}). For example, the binding 
energy of the $S$-wave ground state at 1.2\,$T_c$ is reduced by about 
one order of magnitude, and the state dissolves shortly thereafter, at 
$\sim$1.3\,$T_c$ (Fig.~\ref{fig_PS-F}). The $P$-wave states have 
disappeared already just above $T_c$. 
At the same time the value of the potential at infinity provides a smaller 
selfenergy correction (see Fig.~\ref{fig_C2-Vinf}) leading to a smaller 
effective quark mass and, consequently, a lowered continuum threshold 
compared to using $U$. This, in particular, entails no or only a small rise 
in the in-medium mass of the $J/\psi$ above $T_c$. For the BbS scheme the 
drop in effective mass and the reduction in binding nearly compensate 
each other leading to a stable $J/\psi$ mass until dissolution. For the 
Th scheme the smaller bare-quark mass even leads to a net decrease of the
in-medium $J/\psi$ mass. The euclidean correlator ratios are again very 
similar for the different potentials but exhibit a significantly different 
$\tau$ dependence for the two reduction schemes. For the BbS scheme the 
deviation relative to the vacuum correlator is up to $\sim$50\% while for 
the Th scheme it is no more than 10\%. However, for both schemes the 
temperature evolution is remarkably stable, with variations of no more than 15\% even in the
BbS scheme. Thus the rather large overall deviation originates from the 
reconstructed (vacuum) correlator, where the problem can be traced back 
to the large bare-quark mass which is required in this scheme due to the 
large vacuum binding energy.

\begin{figure*}[!t]
\includegraphics[scale=0.62]{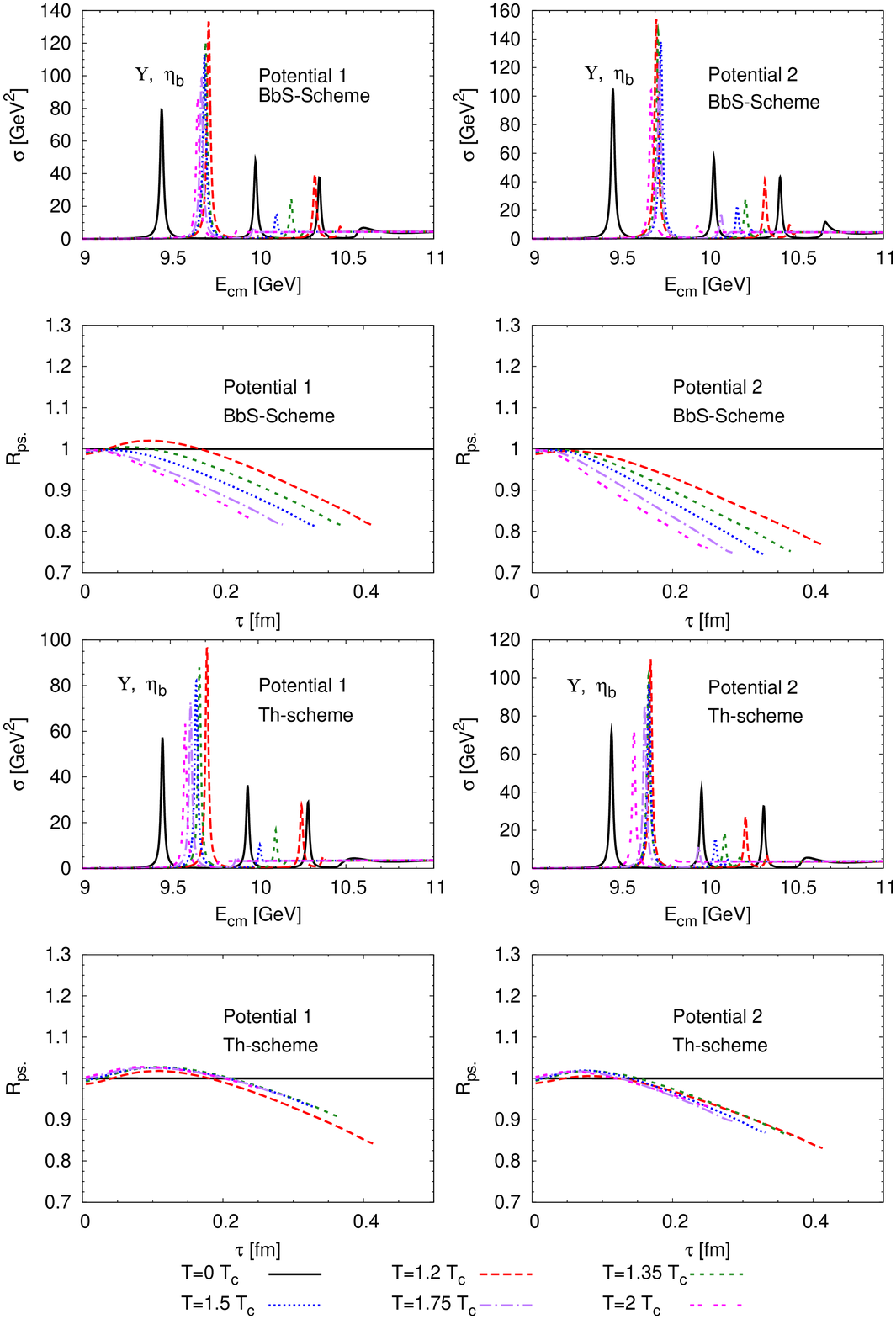}
\caption{(Color online) Bottomonium spectral functions and euclidean-correlator
ratios, using $U$ as potential, for the pseudo-scalar channel at various 
temperatures. We compare BbS- and Th scheme as well as the two different 
potentials.}
\label{PS-nowidth-bottom}
\end{figure*}
\begin{figure*}[!t]
\includegraphics[scale=0.62]{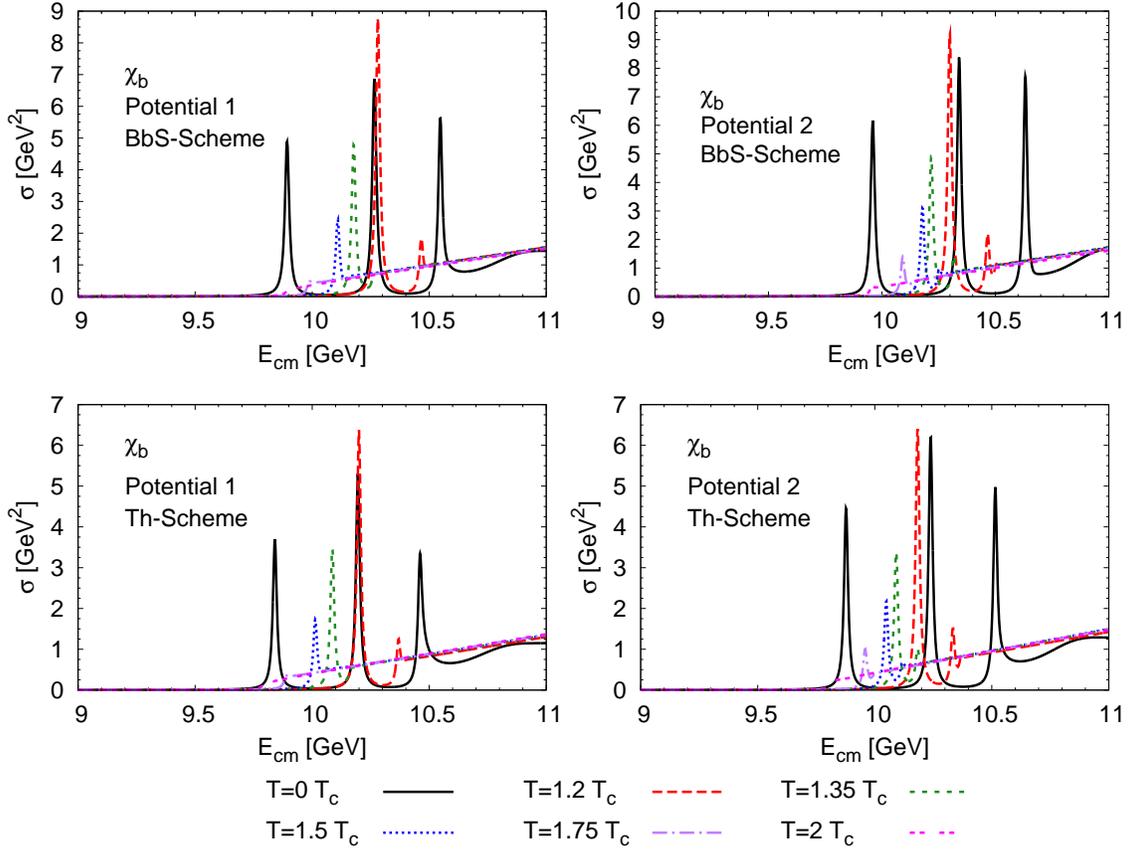}
\caption{(Color online) Bottomonium spectral functions using $U$ as potential 
for the scalar channel at various temperatures using the BbS (upper panels) 
and Th reduction scheme (lower panels).}
\label{S-nowidth-bottom}
\end{figure*}
\begin{figure*}[!t]
\includegraphics[scale=0.62]{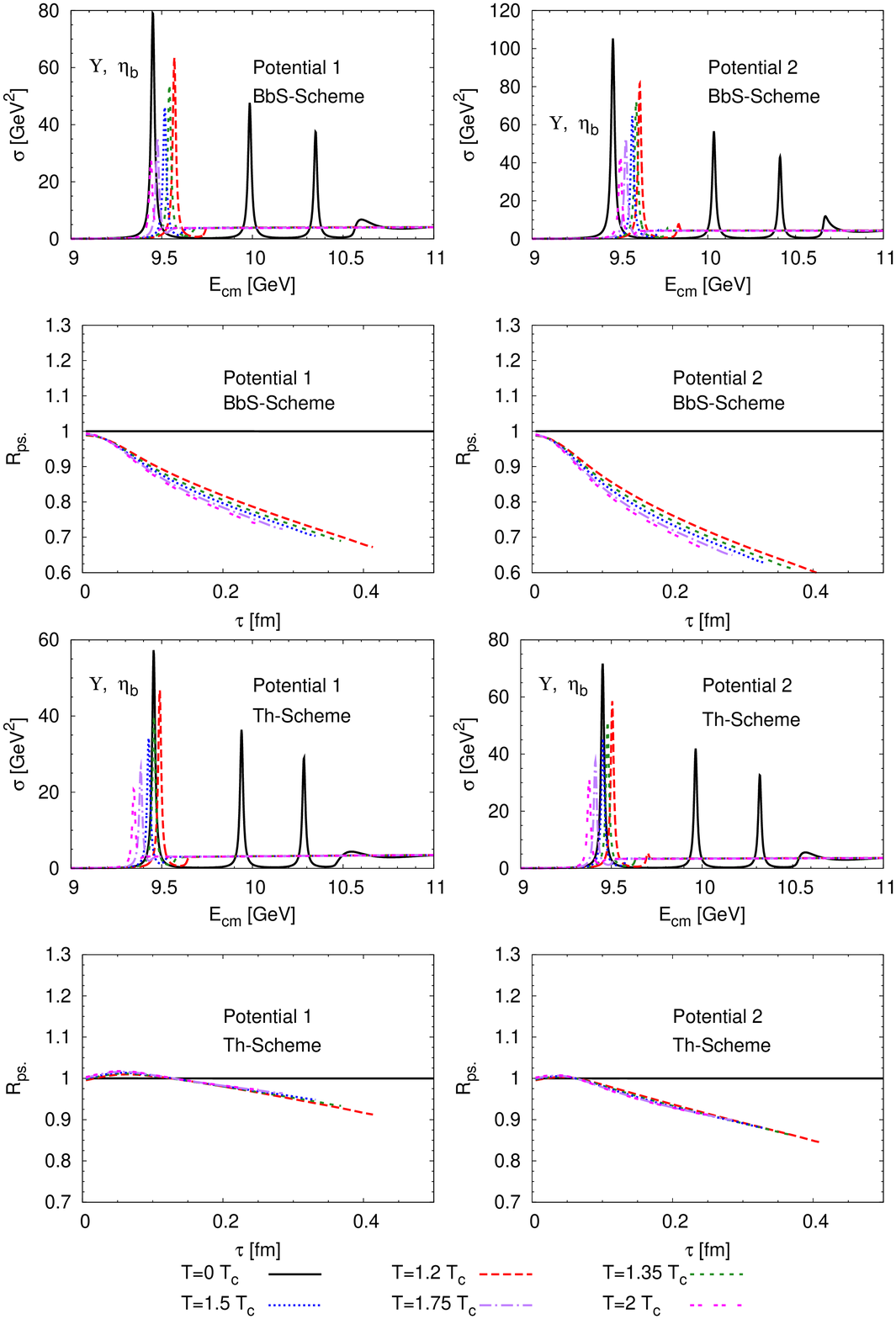}
\caption{(Color online) Same as Fig.~\ref{PS-nowidth-bottom} but 
using $F$ as potential.}
\label{PS-nowidth-F-bottom}
\end{figure*}
\begin{figure*}[!t]
\includegraphics[scale=0.7]{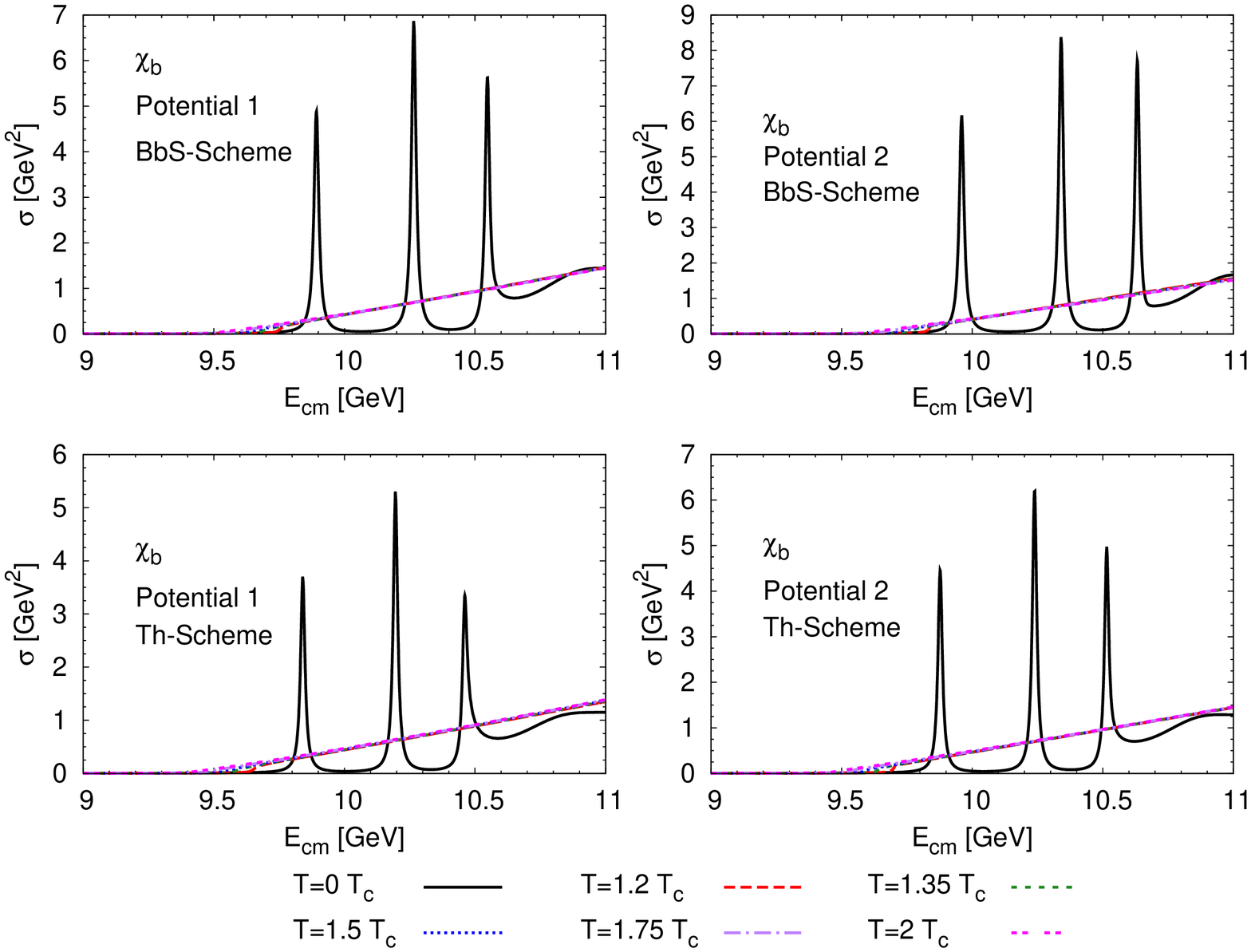}
\caption{(Color online) Same as Fig.~\ref{S-nowidth-bottom} but 
using $F$ as potential.}
\label{S-nowidth-F-bottom}
\end{figure*}

To further map out uncertainties we consider the influence of
a quark width on the correlator ratios.
In Refs.~\cite{Mannarelli:2005pz,vanHees:2007me} it has been found
that the charm-quark width above $T_c$ may be as large as 0.2~GeV.
We injected into Eqs.~(\ref{G2}) a value of $\Gamma_Q=0.1$~GeV for 
the HQ width and plot the results, using $U$ as potential, in 
Fig.~\ref{fig_PS-U-width}.
As an immediate consequence, the $J/\Psi$ width turns out to be about
twice the single-quark quark width, as to be expected.
The ``dissociation" temperature (loosely defined as the temperature 
where the peak height is reduced to less than twice the height of
the continuum) decreases significantly compared to the
narrow-width approximation, to about 1.7\,$T_c$: the broadening of the
resonance structure simply accelerates the merging with the continuum part.
The peak positions (masses) of the narrow-width calculation are basically
preserved. The correlator ratios are increased compared to the calculation
with small quark widths. The magnitude of the effect is relatively small
for the Th scheme where the spread was already small before. For the
BbS scheme the increase is more significantly: the up to 40\% spread
in the narrow-width calculation is reduced to within 30\%.
Similar systematics are also found when using $F$ as potential.

\subsubsection{Bottomonium}
\label{sssec_botttomonium}
In analogy to the charmonium calculations we supply a small 
``numerical width" of 20\,MeV for the $b$ and $\bar b$ quarks. 
The in-medium bottomonium spectral functions and correlator ratios are 
compiled in Figs.~\ref{PS-nowidth-bottom}+\ref{S-nowidth-bottom} for $U$ 
and in Figs.~\ref{PS-nowidth-F-bottom}+\ref{S-nowidth-F-bottom} 
for $F$ as potential.

For the $U$-potential, similar to charmonium, the reduction in binding 
combined with a large effective quark mass (similar as in vacuum) leads to 
an increase in the mass of all bottomonium states right above $T_c$. 
Within the BbS scheme the mass of the lowest 
$\Upsilon$ bound state varies by less than 100\,MeV over the
considered temperature range of 1.2-2\,$T_c$: the lowering of the $b\bar b$
threshold and the loss in binding nearly compensate. But even at 2\,$T_c$ a 
well-defined $\Upsilon(1S)$ bound state persists. The $\Upsilon(2S)$ survives 
up to a temperature of about 1.7\,$T_c$ and shows a much larger variation 
in mass (about 0.5\,GeV) while the $\Upsilon(3S)$ basically dissolves at 
$T_c$. In the Th scheme we observe a similar pattern.
For the euclidean correlator ratios the calculations within the BbS scheme
deviate from one by 20-25\%, more than seen on the lattice. However,  
the relative temperature variations are smaller, ca.~10-15\%. 
In the Th scheme the temperature variations are further reduced to
less than 10\%, and also the deviations from one are smaller, which is 
better in line with the essentially constant lQCD correlator ratios
close to one. Further stabilization of our results is conceivable with 
more realistic in-medium widths and/or improvements in the 
connection between vacuum and in-medium potentials. 
For the $P$-wave $\chi_b$ states the ground state melts at about 
1.7\,$T_c$ while the first excited state dissolves at about 1.2\,$T_c$. 

When using $F$ as potential the reduction in binding is again more 
pronounced, with a dissolution of all excited $S$-wave $\Upsilon$'s and all 
$\chi_b$ states right at $T_c$. Only the $\Upsilon$ ground state survives 
until somewhat above 2\,$T_c$. Compared to the calculation with $U$ as 
potential the strength of the state at 2\,$T_c$ is reduced by a factor of 
$\sim$3, indicating the lower binding, while its mass is about 200\,MeV 
smaller (the loss in HQ mass overcompensates the loss in binding energy). 
Also here the temperature dependence of the ground-state mass is rather 
stable. 
As before, the euclidean-correlator ratios are rather sensitive to the 
interplay of HQ mass, quarkonium binding and the ``polestrength" of
the states. 
The BbS scheme again shows appreciable deviations from one for both
potentials, up to 40\%, while for the Th scheme these are 10-15\%.
However, the spread in the temperature dependence is less than 10\%
for both reduction schemes and lattice inputs.    
We have verified that the inclusion of larger HQ widths has effects 
similar as in the charmonium case, increasing the correlator ratios 
by up to 0.1 units at large $\tau$.

\subsubsection{Discussion of Quarkonium Results}
\label{sssec_disc}
Let us try to summarize and evaluate the findings in the quarkonium sector. 
Within the Th scheme, all $S$-wave correlator ratios (for both lQCD inputs,
for $U$ and $F$, as well as for charmonium and bottomonium) are within 
ca.~15\% of one, for all temperatures between 1.2 and 2\,$T_c$.
For a given calculation (scenario) the {\em relative} deviations within 
this temperature range are, in most cases, even smaller, suggesting that 
the reconstructed correlators play a non-negligible role in the absolute 
uncertainty (e.g., ``residual" hadronic interactions between $D$ and 
$\bar D$ states in the continuum are not accounted for in a single-channel 
$T$-matrix as employed here). 
Within the BbS scheme, we generally find larger deviations of the 
correlator ratios from one (by up to $\sim$50\%); within a given scenario, 
the temperature variations are significantly smaller, up to 30\% (or
even less especially for the free energy). While this may 
overestimate the uncertainty associated with the 4D$\to$3D reduction 
scheme (recall that the BbS scheme has a tendency for over-binding, even 
in vacuum, see also the discussion in Appendix~\ref{app_red}), it 
stipulates that the static approximation (especially for charmonia) 
requires further scrutiny if one aims at an absolute accuracy at the 
10\% level (applications based on the (nonrelativistic) Schr\"odinger 
equation are expected to be beset with larger uncertainty). 
We also corroborated indications found in Ref.~\cite{Cabrera:2006wh}
that effects of a finite spectral width are not negligible either, 
increasing correlator ratios at the 5-10\% level. Our schematic 
implementation of the in-medium widths has only scratched the surface 
of a full many-body calculation utilizing microscopic single-quark 
spectral functions in the $T$-matrix equation (see. e.g., 
Ref.~\cite{Beraudo:2009zz} for a recent perturbative calculation of 
the HQ spectral function in the QGP).    

Our analysis corroborates indications from earlier 
studies~\cite{Wong:2006bx,Cabrera:2006wh,Alberico:2006vw,Mocsy:2007yj} 
that there is currently no decisive discrimination power between the 
different scenarios realized by the use of $U$ (``strong binding") and 
$F$ (``weak binding"). When employing $U$ the mechanism underlying a 
constant (or temperature-stable) correlator
ratio is rather involved, being a combination of 4 components:
On the one hand, the binding energies close to (but above) $T_c$ are rather 
large (several 100\,MeV), together with a large polestrength (due to the 
steepness of the $U$-potential at intermediate distances). On the other hand,
the effective HQ mass, 
governed by $U_\infty(T)$, and thus the $Q\bar Q$ threshold energy, are also
large (basically as in vacuum). With increasing temperature, the binding and 
the polestrength drop, as do the HQ mass and continuum threshold,
thus balancing the (low-energy) strength in the spectral function. 
On the contrary, when employing $F$, the binding already vanishes 
just above $T_c$, and the balance in the spectral function upon 
increasing $T$ is between a further loss of strength in the threshold 
state (cusp) and a reduction 
in the HQ threshold. In particular, with the $F$-potential one does not 
encounter a regime above $T_c$ with a large variation in binding energy. 
However, going further down in temperature, such as regime must inevitably 
occur when approaching the vacuum limit, and similar ``complications" as in 
the $U$-potential calculations above $T_c$ are to be expected. Thus, a 
sensitive test of whether the $F$-potential can be consistent with lQCD 
correlators is in the temperature regime where the largest variation
in binding occurs (which is apparently at or below $T_c$).

\subsection{Heavy-Quark Diffusion in the Quark-Gluon Plasma}
\label{ssec_HQ-diff}
%
\begin{figure*}[t]
\includegraphics[scale=0.62]{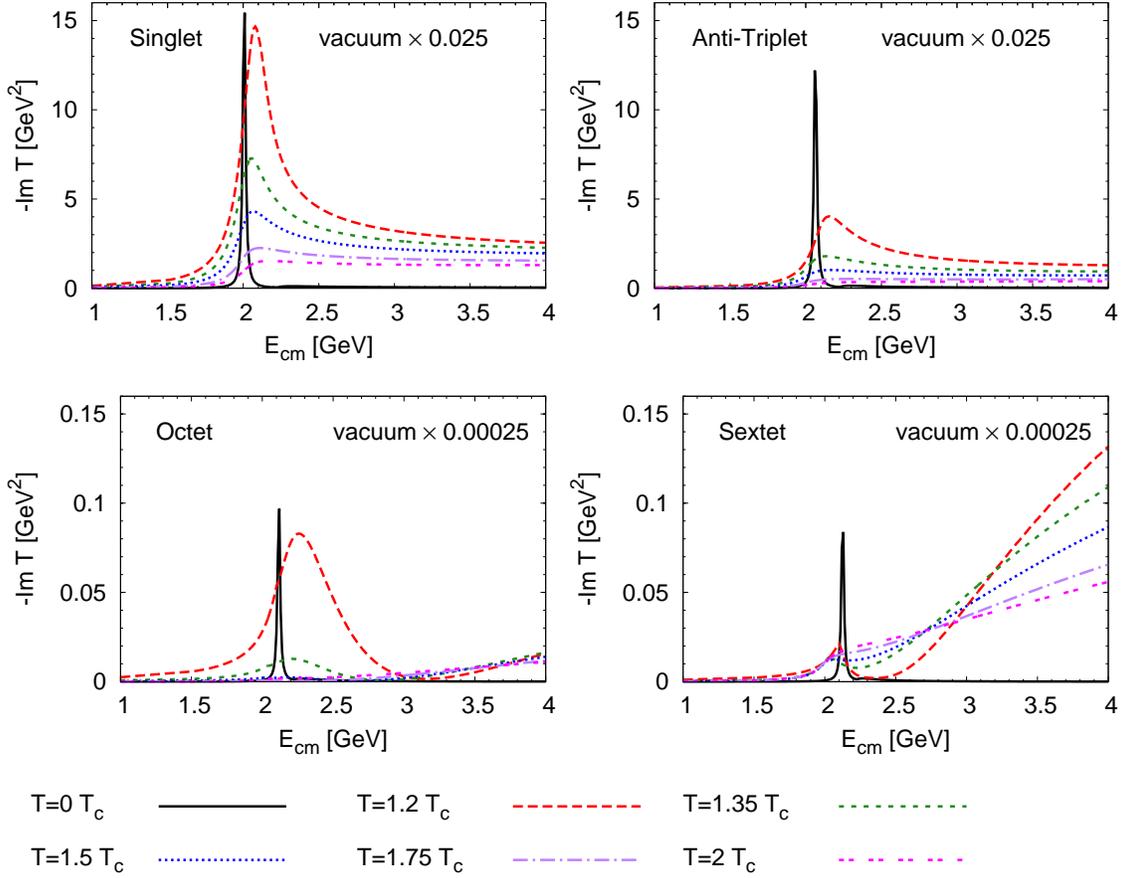}
\caption{(Color online) Imaginary part of the in-medium on-shell 
$T$-Matrix for charm-light quark scattering in the color-singlet (upper 
left), anti-triplet (upper right), octet (lower left) and sextet (lower 
right) channels. In all cases lQCD potential-1 is used for $U$ within the
Thompson scheme. Note the factor 100 difference in the $y$-scales
of the upper and lower panels.}
\label{ImTcq-Th1-width}
\end{figure*}
\begin{figure*}[t]
\includegraphics[scale=0.62]{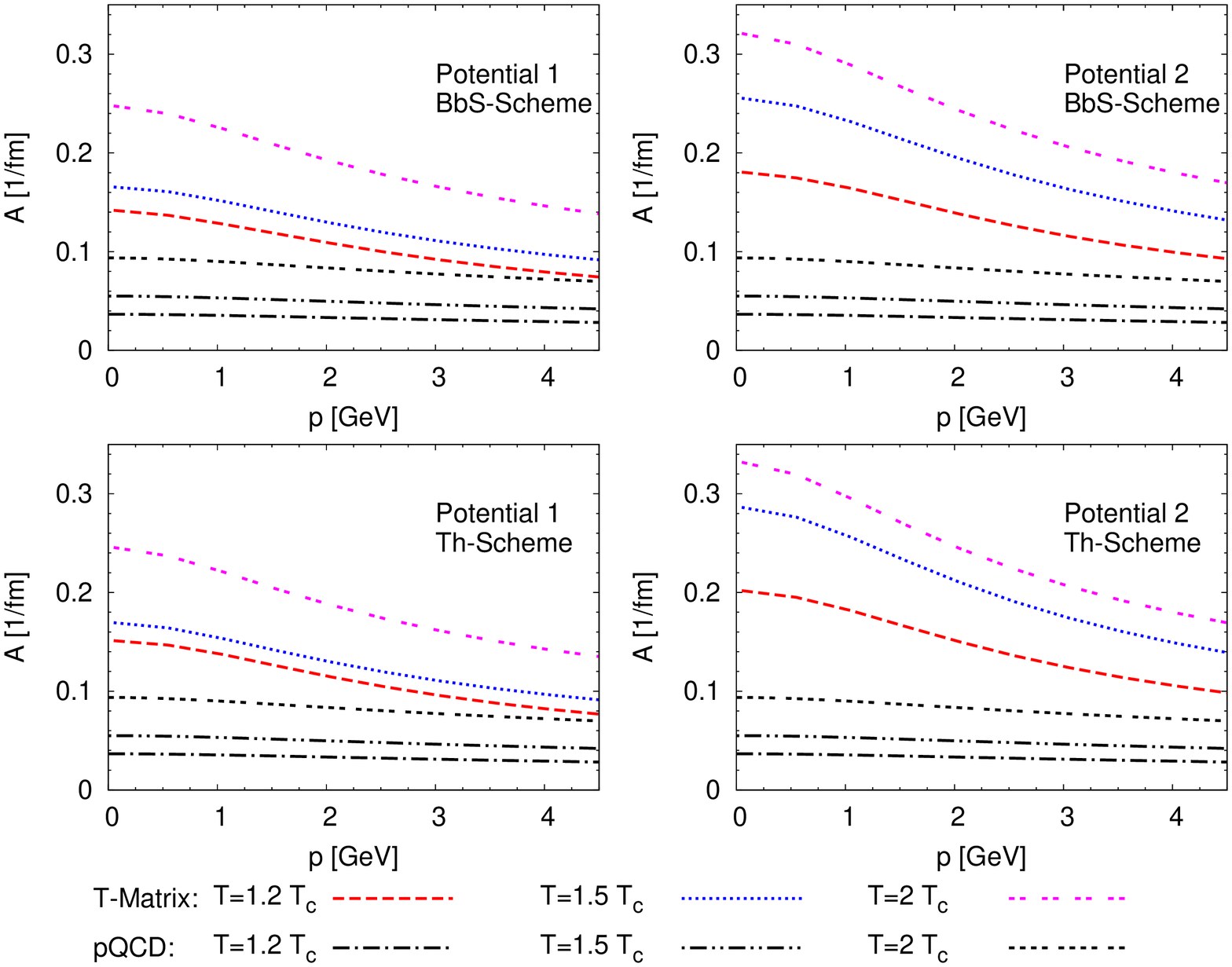}
\caption{(Color online) Charm-quark relaxation rate as a function
of 3-momentum calculated in
the $T$-Matrix approach using $U$ as potential, compared the LO pQCD
with $\alpha_s$=0.4. A perturbative gluon contribution has been added
to the heavy-light $T$-matrix rates.}
\label{A-width-comp-pQCD}
\end{figure*}
\begin{figure*}[t]
\includegraphics[scale=0.62]{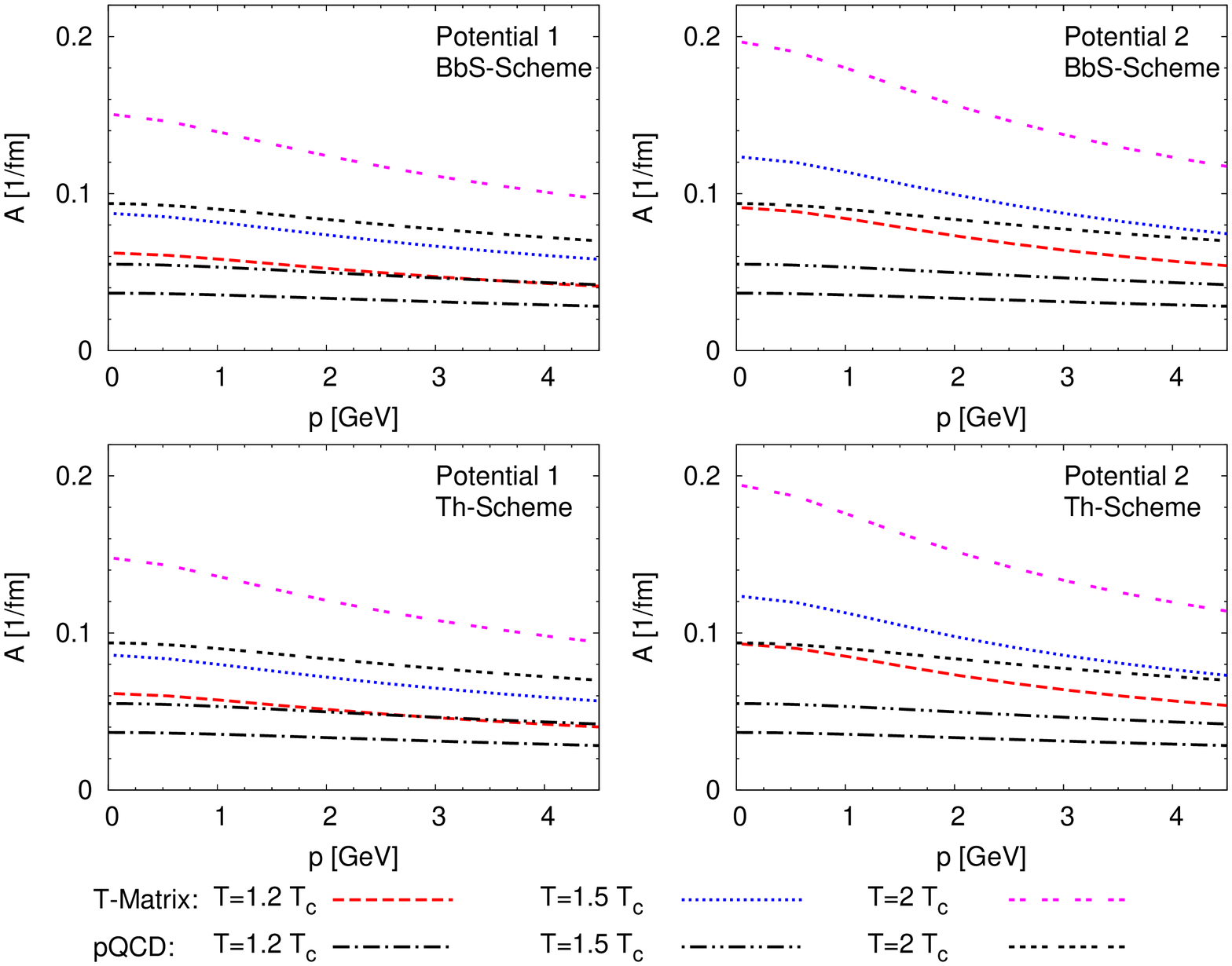}
\caption{(Color online) Same as Fig. \ref{A-width-comp-pQCD} but using 
$F$ as potential.}
\label{A-width-F-comp-pQCD}
\end{figure*}
Following the analysis of in-medium quarkonia we now turn to 
evaluating heavy-flavor transport in the QGP. In the vacuum we have
found that the low-lying $D$-meson spectrum is reasonably well 
reproduced, but also that shallow bound states might occur in colored 
heavy-light two-body channels (recall Fig.~\ref{fig_HLvac}). 
The calculation of the heavy-light $T$-matrix in the QGP requires
an additional input in terms of the in-medium light quark masses (recall 
that the in-medium HQ selfenergy is determined by the infinite-distance 
of the free/internal energy according to Eq.~(\ref{m-eff})). 
Due to chiral symmetry restoration, the vacuum constituent-quark mass 
is expected to approach zero; however, the light quarks and gluons
most likely acquire (chirally symmetric) thermal masses. We 
approximate these by adopting the functional form expected from 
perturbative QCD~\cite{LeBellac}, 
\begin{eqnarray}
m_{\rm th}(T)=\sqrt{\frac{1}{3}}\,g\,T \ , \quad 
m_q(T)=\sqrt{m^2_{q,0}+m_{\rm th}^2(T)} 
\nonumber\\
m_{u,0}= m_{d,0} = 0 \ , \quad m_{s,0}=0.11\,\text{GeV}  \ .
\label{mq-T}
\end{eqnarray}
When implemented into a quasiparticle (QP) description of the QGP 
this form allows to recover an energy density, $\epsilon_{\rm QP}$,
which is roughly 10-20\% below the perturbative value, independent of 
temperature~\cite{Levai:1998xx,Peshier:2002ww}. We fix the strong 
coupling in Eq.~(\ref{mq-T}) at $g=2.3$, resulting in 
$\epsilon_{\rm QP}/\epsilon_{\rm SB}\simeq0.83$, consistent with recent
lQCD calculations~\cite{Cheng:2009zi} for $T\ge1.4\,T_c$. This value
for $g$ is also compatible with our perturbative calculations for 
scattering off thermal partons ($\alpha_s\simeq$0.4).

In Fig.~\ref{ImTcq-Th1-width} we compile in-medium charm-light 
$T$-matrices using $U$ for potential-1 within the Th scheme, with an 
in-medium single-quark width of 100\,MeV (uncertainties due to 
reduction scheme and potential are exhibited in the context of the
thermal relaxation rates below). 
In the medium the color-sextet and -octet correlations fade rapidly 
due to screening of the attractive string part of potential 
(recall Fig.~\ref{fig_VT}). The meson 
(color-singlet) and diquark (color anti-triplet) channels feature 
broad ``Feshbach resonances" (i.e., resonances at threshold) up to 
$\sim$1.5\,$T_c$. Compared to previous $T$-matrix 
results~\cite{vanHees:2007me}, the diquark state is slightly
more robust, due to the refined (color-blind) treatment of the string 
term (e.g., the ratio of peak heights for color anti-triplet to color 
singlet at 1.2~$T_c$ is about twice as large). 
For the charm-strange correlations similar patterns are found.

\begin{figure*}[t]
\includegraphics[scale=1.0]{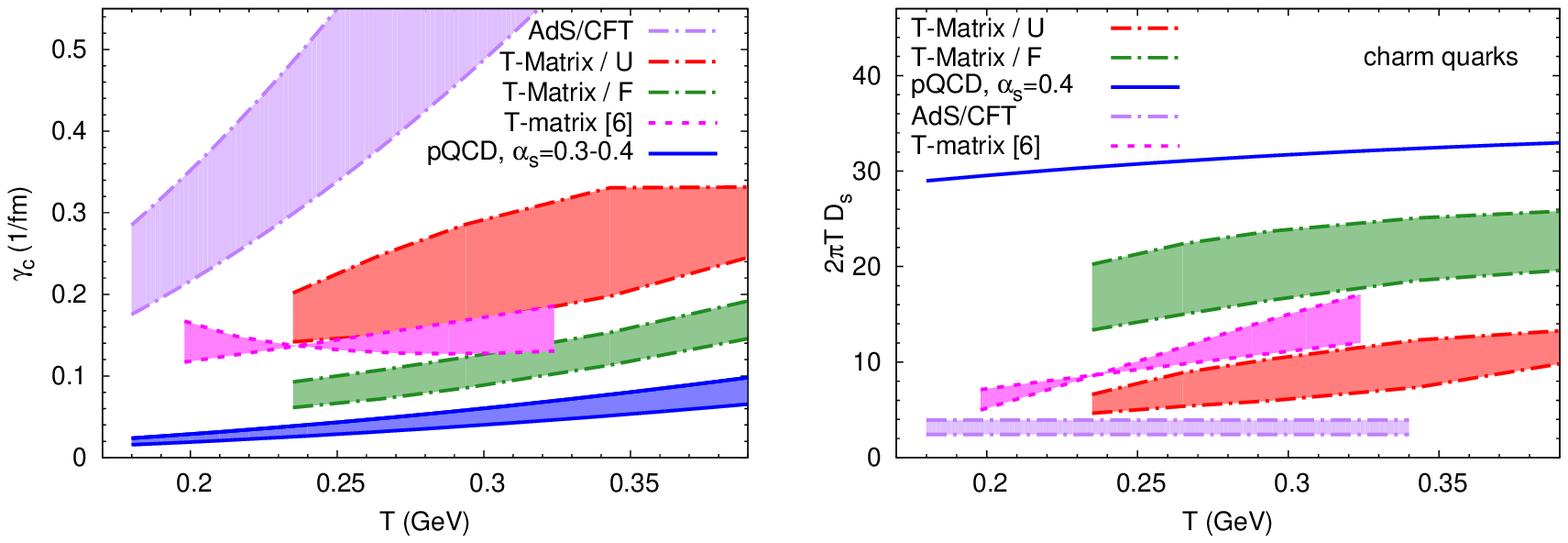}
\caption{(Color online) Comparison of our results for charm-quark relaxation
rates (left plot) using $U$- and $F$-potentials to previous $T$-Matrix 
calculations~\cite{vanHees:2007me} and estimates from 
AdS/CFT~\cite{Akamatsu:2008ge,Gubser:2006qh}. The $T$-Matrix rates have been
augmented by perturbative scattering off thermal gluons. The right panel 
shows the corresponding spatial diffusion constants.}
\label{gamma-vsT-comp}
\end{figure*}
Next we calculate HQ relaxation rates. The original suggestion of 
nonperturbative
effects in HQ diffusion has been put forward in Ref.~\cite{vanHees:2004gq} 
using an effective resonance model where the masses and coupling strengths
were free parameters; within reasonable ranges of these, a factor 2-4 
shorter thermalization times compared to pQCD were found. Subsequently,   
heavy-light $T$-Matrix calculations~\cite{vanHees:2007me} were carried out 
to render the schematic estimates more quantitative (and to check for the
existence of $D$-meson resonances in the QGP), roughly confirming the results
of the resonance model {\em if} the $U$-potential is employed. Here, we 
elaborate for the first time a quantitative connection to in-medium 
quarkonium properties.
With the potential and all other parameters determined our relaxation 
rates, $A(p)$, are predictions of the approach. They are calculated utilizing 
Eq.~(\ref{Ap}) and displayed in Figs.~\ref{A-width-comp-pQCD} and 
\ref{A-width-F-comp-pQCD} as a function of the HQ momentum for several
temperatures above $T_c$. For completeness, we have added to the
$T$-matrix results the contribution from HQ scattering off gluons using 
LO pQCD diagrams (including Debye-screening) with a coupling constant 
$\alpha_s$=0.4. At the lowest temperature, $T=1.2$\,$T_c$, we find  
$\gamma_c=0.14-0.2$\,fm$^{-1}$, where most of the variation is due to 
the potential choice while the reduction schemes 
agree within 10\% for a given potential (pQCD scattering off gluons 
contributes ca.~0.025\,fm$^{-1}$). Thus, in the scattering regime the 
dependence on the reduction scheme is less pronounced than for bound 
states (see also Appendix~\ref{app_red}). 
The relaxation rate increases to 0.25-0.33\,fm$^{-1}$ at 2\,$T_c$ 
(again with most of the spread owing to the difference in the potentials;
pQCD scattering off gluons contributes ca.~0.07\,fm$^{-1}$). 
The magnitude of the low-momentum relaxation rates at $T=1.2$\,$T_c$ 
(2\,$T_c$) is a factor 4-5 (2.5-3.5) larger than for a LO pQCD 
calculation for scattering off thermal quarks, antiquarks and gluons 
with $\alpha_s$=0.4. They are slightly larger than the previous 
$T$-matrix results of Ref.~\cite{vanHees:2007me}, 
where $\gamma_c$=0.12-0.19\,fm$^{-1}$ has been obtained over the
temperature range $T=$~1.1-1.8\,$T_c$ for parametrizations of yet two 
other (quenched~\cite{Kaczmarek:2003dp,Wong:2004zr} and 
$N_f$=2~\cite{Kaczmarek:2003ph,Shuryak:2004tx}) lQCD-based internal 
energies. In the previous calculations~\cite{vanHees:2007me} a 
constant charm-quark mass of $m_c=1.5$\,GeV was used while we here 
include the in-medium selfenergy from the infinite-distance limit of 
the internal (or free) energy. When using $U$, $m_c^*$ is larger than
1.5\,GeV up to temperatures of ca. 1.9\,$T_c$ (potential-1 with Th 
scheme, see Tab.~\ref{tab_mass} and right panel of Fig.~\ref{fig_C2-Vinf}).
The extra interaction strength in our present calculation compared
to Ref.~\cite{vanHees:2007me} is mostly due to the color-blind
treatment of the string term, particularly in the diquark channel.  
\begin{figure*}[t]
\includegraphics[scale=0.62]{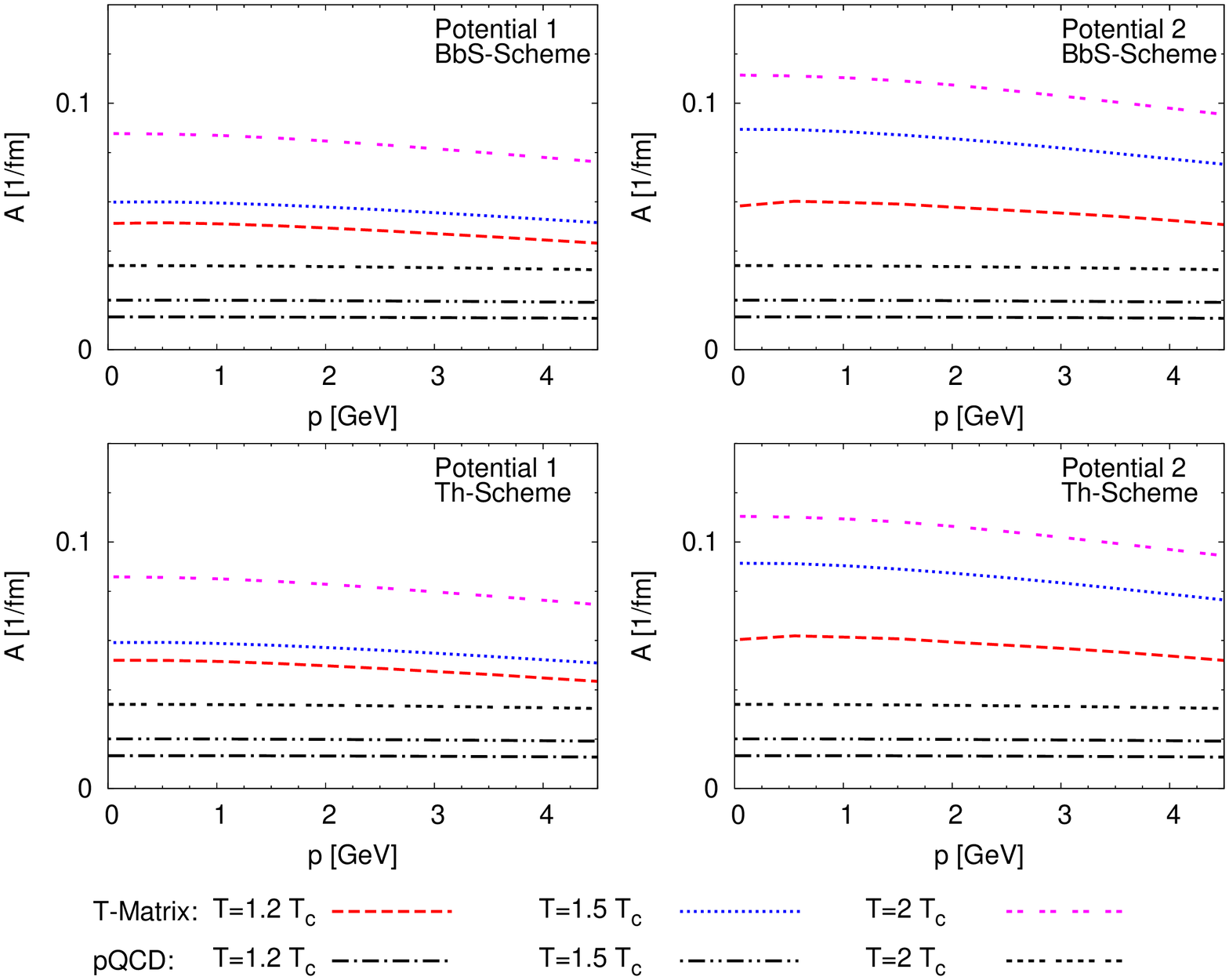}
\caption{(Color online) Bottom-quark relaxation rates as a function
of 3-momentum calculated in the $T$-Matrix approach using $U$ as potential 
(plus perturbative scattering off thermal gluons), compared to LO pQCD.} 
\label{A-width-comp-pQCD-bottom}
\end{figure*}
\begin{figure*}[t]
\includegraphics[scale=0.62]{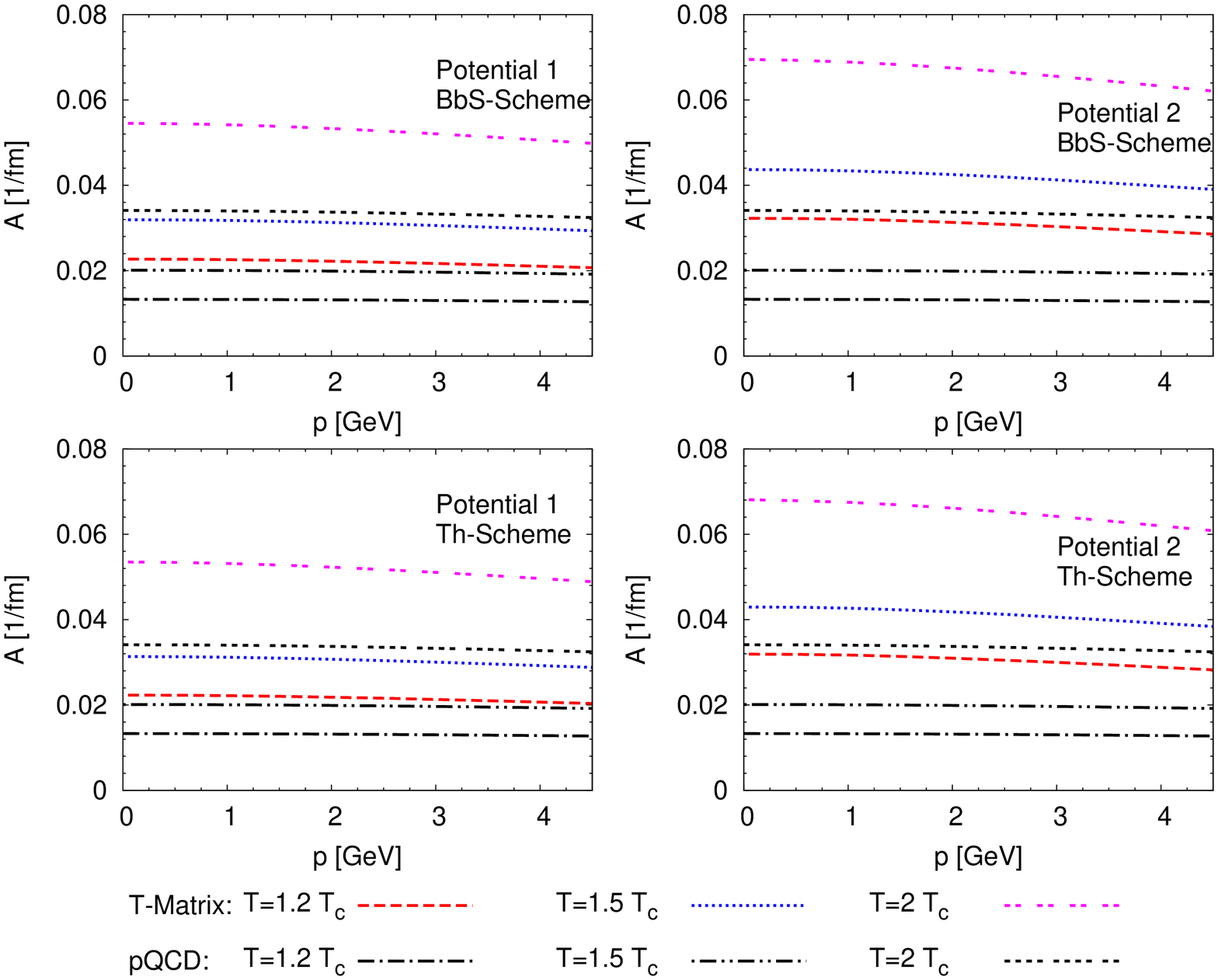}
\caption{(Color online) Same as Fig. \ref{A-width-comp-pQCD-bottom} 
but using $F$ as potential.}
\label{A-width-F-comp-pQCD-bottom}
\end{figure*}
We emphasize that the in-medium HQ masses as used here are
mandatory to maintain consistency with the quarkonium
correlator ratios where they play a critical role in balancing the 
changes in binding energy. 
Our investigations actually show that the 
internal energy based on the quenched lQCD input from 
Refs.~\cite{Kaczmarek:2003dp,Wong:2004zr} leads to euclidean 
correlator ratios for quarkonia which exhibit a large 
temperature variation (decrease with increasing $T$) incompatible 
with lQCD results, i.e., well beyond the 30\% error margin deduced in 
Sec.~\ref{sssec_charmonium}. The large temperature variation (screening)
in the underlying potential leads to a {\em decrease} of the thermalization
rate with temperature. This feature is not confirmed in the more
quantitative calculations presented here. However, the increase with 
temperature of $\gamma_c$ for our $T$-matrix (plus pQCD gluon 
scattering) calculations is significantly slower than for the LO pQCD 
calculations with temperature-dependent Debye mass: for $T=1.2\to2T_c$ 
the former increase by a factor of $\sim$1.7 (less for the $T$-matrix
contribution alone), compared to a factor of $\sim$2.5 for LO pQCD only 
(light anti-/quarks and gluons). Furthermore, in the $T$-matrix 
calculations $A(p)$ decreases appreciably with increasing 3-momentum 
while the LO pQCD results are almost constant. This is simply due to the 
fact that with increasing 3-momentum the charm quark is less likely to 
excite a low-energy Feshbach resonance in collisions with thermal quarks
or antiquarks. At high 3-momentum, resummation effects in the $T$-matrix 
cease and the relaxation rates come closer to the LO pQCD results (recall
the importance of the proper relativistic factors for this behavior).
The difference at high 3-momentum is mostly due to the smaller value of 
the screening mass of the Coulomb term in our lQCD fit relative to the 
pQCD value, $m_D^{\rm pQCD}=\sqrt{1+N_c/6}\,gT$.
As in Ref.~\cite{vanHees:2007me}, the dominant contribution to the HQ
relaxation rate originates from the $S$-wave meson (color-singlet) meson 
and diquark (color-triplet) channels, while the octet and sextet channels 
are suppressed (even at 1.2\,$T_c$), as is immediately inferred from the 
magnitudes of the corresponding $T$-matrices in Fig.~\ref{ImTcq-Th1-width}.
The $P$-wave channels contribute about 30\% of the $S$-waves. 

When using $F$ instead of $U$ as potential the low-momentum charm-quark
relaxation rate is reduced by approximately a factor of $\sim$2, but still 
larger by a factor of $\sim$2 than the LO pQCD results, 
cf.~Fig.~\ref{A-width-F-comp-pQCD}. Consequently, they come closer to the
LO pQCD results at high momentum, even though a significant enhancement
persists even at $p$=5\,GeV (mostly due to the differences in screening mass
as mentioned above).

To put our results in context with other approaches we display in 
Fig.~\ref{gamma-vsT-comp} (left panel) the temperature dependence of 
the relaxation rate at zero momentum for different models. Specifically, 
we compare our results for $U$ and $F$ to LO pQCD, to earlier 
$T$-Matrix calculations~\cite{vanHees:2007me} and to estimates from 
gravity-gauge duality (AdS/CFT)~\cite{Akamatsu:2008ge,Gubser:2006qh}
 (see also Refs.~\cite{Peshier:2008bg,Gossiaux:2008jv} 
for LO calculations with running coupling). 
The uncertainty bands associated with our $T$-matrix calculations are 
largely governed by the differences in the underlying lQCD input.    
As discussed above, the results using $U$ overlap with 
the earlier $T$-Matrix calculations (where also $U$ has been used as 
potential), especially when the latter would be calculated with a
color-blind string term. When using $F$ the results are closer to, 
but still significantly above, LO pQCD. The AdS/CFT rates are markedly 
larger than any of the $T$-matrix rates, except for extrapolations close 
to $T_c$. 

In the right panel of Fig.~\ref{gamma-vsT-comp} we compile the 
temperature dependence of the spatial diffusion coefficients, 
\begin{equation}
D_s = \frac{T}{m_c \gamma_c} \ ,
\label{Ds}
\end{equation}
for the above discussed approaches. We plot $D_s$ in units of the 
thermal wave length of the medium, $1/(2\pi T)$, which renders it 
suggestive for a connection to the widely discussed ratio of viscosity 
to entropy-density.  E.g., in kinetic theory for a weakly 
interacting gas one has the approximate relation 
\begin{equation}
\frac{\eta}{s}\approx \frac{1}{5} T D_s \ . 
\end{equation}
In the strongly coupled limit of the AdS/CFT correspondence, one finds
the same parametric dependence, albeit with a different numerical coefficient
(the conjectured lower bound of $\eta/s=1/4\pi$ corresponds to
to diffusion at the thermal wavelength, $D_s\simeq1/(2\pi\,T)$). 
Besides the quantitative comparison of the $D_s$ values their
$T$-dependence is of particular interest. It is constant for AdS/CFT
(which has no no scale other than temperature; note that the HQ mass
is effectively divided out in Eq.~(\ref{Ds})) and almost constant
for LO pQCD and the $T$-matrix approach with $F$ as potential, decreasing
by less than 5\% and up to 30\%, respectively, for $T=2\to1.2\,T_c$. The 
variation is larger, ca.~50\%, if $U$ is used as potential. 
The largest variation of more than 50\% is found with the quenched 
lQCD input~\cite{Kaczmarek:2003dp,Wong:2004zr} for $U$ in
the previous $T$-matrix calculations, but, as we indicated above, this
$T$-dependence is incompatible with the small temperature variation
in the euclidean quarkonium correlator ratios. Nevertheless, our
current, better constrained $T$-matrix calculations support a decreasing
trend when approaching the ``critical" temperature from above, as typical
for many substances at or in the vicinity of a second-order transition.

In Figs.~\ref{A-width-comp-pQCD-bottom} and \ref{A-width-F-comp-pQCD-bottom}
we display the relaxation rates for bottom quarks for the $U$- and 
$F$-potential, respectively. The general trends (and quantitative 
enhancements over LO pQCD) are very similar to the charm case so that 
an analogous discussion applies which we do not reiterate here.  

\section{Summary and Conclusions}
\label{sec_concl}
We have set up a common framework to evaluate properties of open and 
hidden heavy-flavor states in the QGP. A thermodynamic $T$-matrix formalism 
for heavy quarkonia and heavy-light quark interactions has been combined 
with input potentials estimated from heavy-quark free energies computed in 
lattice QCD. Compared to earlier calculations, we have refined this link 
by utilizing a field-theoretic ansatz for an effective in-medium gluon 
propagator. This enabled the fits to be carried out at the level of the 
color-average free energy while disentangling color-Coulomb and confining
interactions and thus gain insights into their medium modifications via
the temperature dependence of the associated fit parameters (screening 
masses and coupling strengths).   
The $T$-matrix calculations further allowed us to identify appropriate
relativistic corrections to the static potential, including differences 
between vector and scalar interactions for the color-Coulomb
and confining parts, respectively. E.g., a color-Breit correction 
naturally emerges for the Coulomb term. The relativistic corrections are 
crucial to establish quantitative consistency for high-energy scattering
between perturbative QCD and the $T$-matrix in Born approximation.
This connection is a prerequisite for a simultaneous treatment of bound 
and scattering states, which was one of the main objectives of our work. 

The bare masses of the charm and bottom quark have been fixed to the 
(spin-averaged) mass of the quarkonium ground states, $\eta_c$-$J/\psi$ 
and $\Upsilon$, in vacuum. The resulting mass splittings for the 
excited states agree with the experimental values within ca.~$\pm$10\%, 
which is smaller than the effects due to hyperfine interactions which 
have been neglected in this work. The largest source of uncertainty
turned out to be the static reduction scheme underlying the scattering
equation, while the 2 considered lattice potentials induced smaller
variations. We also verified that the vacuum
$D$- and $B$-meson states are reasonably well recovered when using
typical values for the constituent light- and strange-quark mass.  
As a by-product, we found that the scalar treatment of the confining
force leads to shallow bound states in the color-sextet and -octet
channels in vacuum, which might be relevant for a rather rich spectroscopy
of narrow four-quark states as discussed in the recent literature.  

Our finite-temperature calculations have been carried out within two 
scenarios of adopting an in-medium potential from the lattice results, 
either the free ($F$) or internal ($U$) energy. First, we calculated  
spectral functions and pertinent euclidean-correlator ratios for heavy 
quarkonia. We confirmed the earlier found trend that for $F$ charmonia 
dissolve rather close to $T_c$ ($T_{diss}$$\simeq$$1.2T_c$) while for 
$U$ the $J/\psi$ may survive up to 2-2.5~$T_c$. However, both scenarios 
can lead to almost constant correlator ratios, and thus to agreement
with lattice QCD results for this quantity. The reason is a small 
in-medium HQ mass correction when using $F$, while it is larger for $U$. 
As in the vacuum, we found significant variations due to the static 
reduction scheme, reflected by deviations of up to $\sim$40\% in the 
correlator ratios at a given temperature. However, within a given 
reduction scheme, potential choice and lattice input, the relative
{\it temperature} variation of the correlator ratios is usually much 
smaller. This suggests that future studies should scrutinize corrections
to the static approximation, but also the role of the reconstructed 
(vacuum) correlator figuring into the denominator of the ratios, 
especially close to threshold where hadronic ($D\bar D$) correlations 
could become important.  

For heavy-flavor transport in the QGP, the use of $U$ leads to a factor 
of $\sim$2 smaller thermalization times and (spatial) diffusion constant 
compared to $F$. This is largely due to ``Feshbach"-type resonances in 
meson and diquark channels up to 1.3-1.5\,$T_c$, but nonperturbative 
rescattering strength persists in the heavy-light $T$-matrix for 
temperatures beyond 2\,$T_c$. Even when using $F$ as potentail, these 
effects lead to up to a factor of 2 faster thermalization compared to 
perturbative scattering. The uncertainty due to the reduction scheme is 
smaller for heavy-quark transport coefficients than for quarkonium 
correlator ratios.
The screening effects in the interaction generate a significant increase 
of the spatial diffusion constant (in units of the thermal wavelength) 
with temperature (especially for $U$), suggestive for a minimum 
toward $T_c$. 

Our analyses suggest that a thermodynamic $T$-matrix approach can be 
used to establish quantitative relations between quarkonium survival and 
heavy-quark transport in the QGP. In particular, we have assessed 
uncertainties associated with commonly applied static (potential) and 
nonrelativistic approximations. While the latter are mandatory in the
scattering regime, the former turned out to be on the few-tens of 
percent level, which is relatively large for the lattice correlator
ratios, but relatively small in the context of current estimates for
heavy-quark diffusion coefficients. A pressing issue remains the additional 
uncertainty in the definition of a finite-temperature potential, especially
when based on model-independent input from thermal lattice QCD. 

Several directions for future investigations emerge from our studies.
As already mentioned, retardation effects and the influence of virtual 
anti-particle contributions need to be addressed, especially in the 
bound-state regime, e.g., by replacing the $T$-matrix by a 
Dyson-Schwinger formalism at finite temperature. Such studies could 
also facilitate the treatment of heavy-quark interactions 
with thermal gluons beyond the perturbative level. 
A more microscopic treatment of the heavy-quark width figuring into
the 2-particle propagator of the scattering equation is desirable and
in principle straightforward. Additional finite-width effects arise  
via inelastic interaction channels, which can be implemented via
coupled channels into the $T$-matrix equation. For example,
gluon radiation is expected to become important for high-energy
charm-quark scattering and/or quarkonium dissolution, while $D\bar D$ 
or even magnetic charge-anticharge states could improve the description 
around $T_c$ and extend it to temperatures below $T_c$.
Heavy-quark susceptibilities, or more generally correlators of charm 
quarks with conserved charges (e.g., baryon or strangeness), which 
are computed with good accuracy in thermal lattice QCD, can be 
calculated with our $T$-matrix. Here, the presence of broad resonances 
does not necessarily imply large signals in such quantities. 
Finally, the in-medium quarkonium and heavy-quark transport properties
should be implemented into a comprehensive phenomenological analysis
of pertinent observables in heavy-ion collisions, e.g., via rate
equations and/or Langevin simulations in a realistic bulk medium
evolution. This will provide quantitative tests of the equilibrium 
results in current and future experiments and thus advance our 
understanding of strongly coupled QCD matter at temperatures around 
and above $T_c$. Work along some of these lines has been initiated. 

\vspace{0.5cm}

{\bf Acknowledgments} \\
We gratefully acknowledge discussions with D.~Cabrera, H.~van Hees, 
O. Kaczmarek, 
T.-S.H.~Lee, R. Machleidt, M. Mannarelli, P. Petreczky and  A. Vairo
on various aspects of this work. We especially thank 
O.~Kaczmarek an P.~Petreczky for providing their lattice QCD results.
This work is supported by the U.S. NSF under grant numbers PHY-0449489 
(CAREER) and PHY-0969394, and by the Alexander-von-Humboldt Foundation.

\begin{appendix}
\section{Differences in the reduction scheme}
\label{app_red}
\begin{figure}[t]
\includegraphics[scale=0.62]{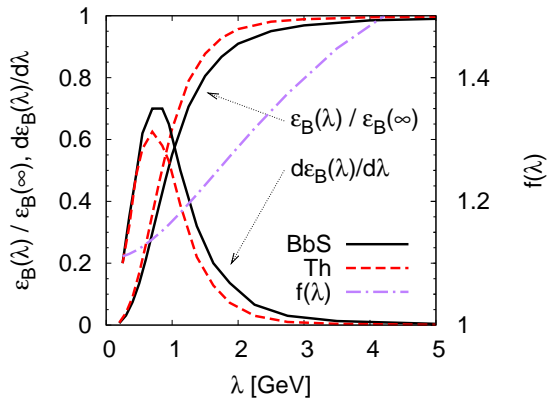}
\caption{(Color online) Dependence of the vacuum $J/\Psi$ binding
energy on a momentum cutoff introduced in the $T$-matrix integral
in Eq.~(\ref{Tmat}).}
\label{fig_tmat-intgd}
\end{figure}
In this appendix we discuss differences between the BbS and Th reduction 
scheme. For this purpose we concentrate on the case of heavy
quarkonium bound states. From Eqs.~(\ref{G2}) it follows that the
BbS and Th 2-particle propagators differ as
\begin{eqnarray}
G^{\rm BbS}_{12}(E,k)=
\frac{4\,\omega_{Q}(k)}{E+2\,\omega_{Q}(k)}\,G^{\rm Th}_{12}(E,k) \ 
\end{eqnarray}
due to a different treatment of the left-hand cut (virtual antiparticle 
contributions). Both reduction schemes should give very similar results for 
the $T$-Matrix if 
\begin{eqnarray}
f(k)=\frac{4\,\omega_{Q}(k)}{E+2\,\omega_{Q}(k)}\approx 1 \ .
\end{eqnarray}
This condition is rather well satisfied in the scattering region,
i.e., above the 2-particle threshold, where the integral is dominated by 
the pole (unitarity cut) of the propagator, $E-2\,\omega_{Q}(k)\approx 0$, 
which implies $E \approx 2\,\omega_{Q}(k)$. However, in the bound-state 
regime, i.e.,  below threshold, the situation can be different.
For example, in the extreme case of $E\to0$ the difference between the 
propagators becomes as large as a factor of 2, entailing large
discrepancies in the results for the $T$-Matrix. 
Let us try to asses the differences more quantitatively for the case at 
hand, i.e., for the binding energy of the charmonium ground state in 
vacuum. Our results for the BbS and Th scheme show a ca.~25\% difference
in the $J/\Psi$ binding energy (the explicit values are quoted in the 
legend of Fig.~\ref{fig_eps-psi-U}). As a rough guideline, the influence 
of $G$ on the binding may be estimated by formally writing the 
solution of the $T$-matrix as $T=V/(1-GV)$. At the bound-state energy, 
one has $GV=1$, and thus a 25\% change in $G$ approximately ``mimics" 
a 25\% stronger potential, or binding energy (for the same static input 
potential). Thus, for the BbS propagator in the $T$-matrix integral one 
should expect, on average,   
\begin{eqnarray}
f(k)=\frac{4\,\omega_{Q}(k)}{E+2\,\omega_{Q}(k)}\approx 1.25 \ .
\end{eqnarray}
To estimate the relevance of the integration momenta we apply a cutoff,
$\lambda$,  in the $T$-matrix equation (\ref{Tmat}) and study the
dependence of $f(\lambda)$ and the $J/\psi$ binding energy on this cutoff,
as displayed in Fig.~\ref{fig_tmat-intgd}. Taking as an approximative representative 
momentum the one by which half of the binding is built up 
($\lambda\simeq 1$\,GeV) and evaluating the ``BbS factor" at this value, 
one finds $f(\lambda)\simeq1.2$. 
The magnitude of the deviations between BbS and Th for bound states can 
thus be roughly accounted for and is expected to become larger with 
increasing ratio of binding energy to the mass of the constituents.

%
\section{Binding energies}
\label{app_Ebind}
%
%
\begin{figure}[t]
\includegraphics[scale=0.62]{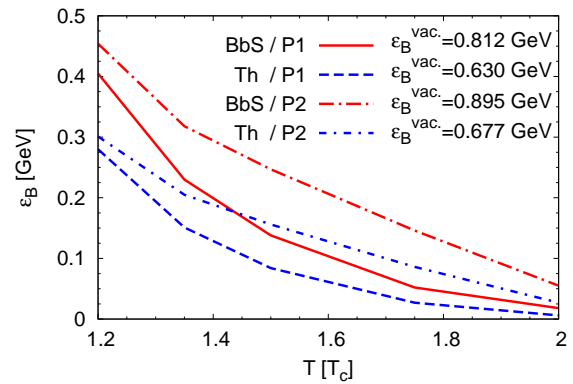}
\caption{(Color online) Temperature dependence of the binding energy
($\epsilon_B$) of the $J/\Psi$ (or $\eta_c$) using $U$ as potential,
for various combinations of reduction scheme and lQCD input.}
\label{fig_eps-psi-U}
\end{figure}
%
\begin{figure*}[t]
\includegraphics[scale=1.0]{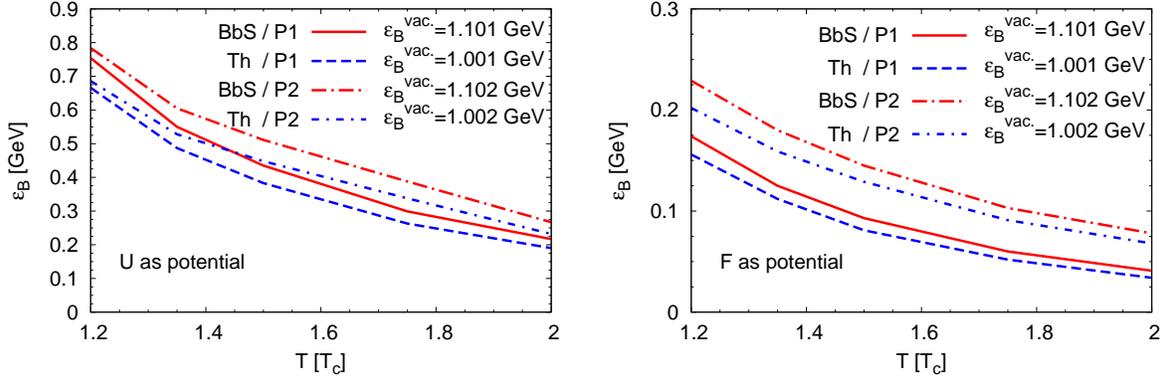}
\caption{(Color online) Temperature dependence of the binding energy
($\epsilon_B$) of the $\Upsilon$ (or $\eta_b$) using $U$ as potential (left plot) and $F$ as potential (right plot),
for various combinations of reduction scheme and lQCD input.}
\label{fig_eps-ups-U-F}
\end{figure*}
In this appendix we compile the temperature dependence of the binding 
energies of the $J/\psi$ and $\Upsilon$ ground states. We define 
the binding energy as the difference between the quark-antiquark
threshold, 2$m_Q$, and the mass of the state in question. In the 
vacuum the $J/\psi$ ($\eta_c$) binding energy is about 0.65(0.85)\,GeV 
for the Th (BbS) scheme. The in-medium binding energies are shown in 
Fig.~\ref{fig_eps-psi-U} when using $U$ as potential
(for $F$ the state already dissolves at about 1.3 $T_c$). 
One observes that the BbS scheme leads to a steeper dependence of the 
binding on temperature compared to the Th scheme
while the melting temperature is quite similar in both cases.
 \begin{figure*}[t]
\includegraphics[scale=0.62]{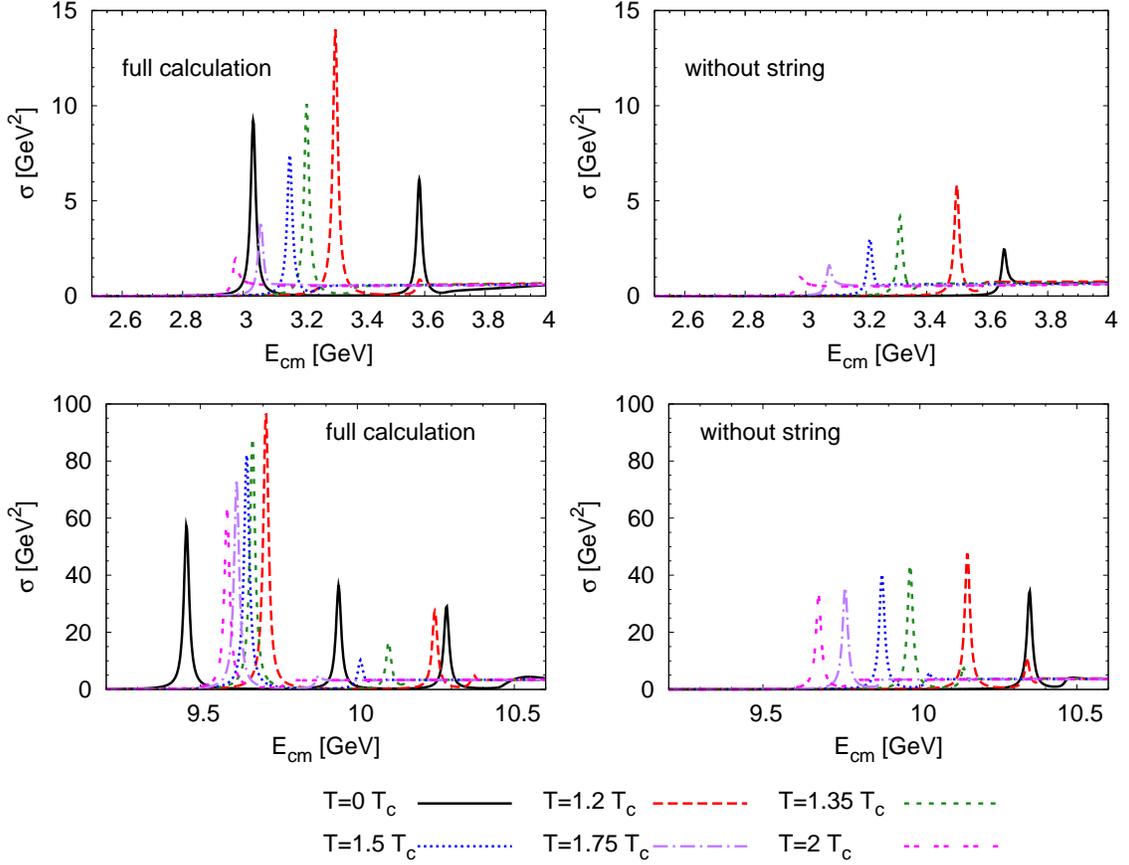}
\caption{(Color online) Comparison of the pseudoscalar spectral functions 
for charmonium (upper panels) and bottomonium (lower panels) using the full 
potential (left column) and the color-Coulomb term only (right column). 
In all cases $U$ has been used as potential together with the Th reduction.}
\label{Spec-nostring}
\end{figure*}

A similar pattern occurs for the $\Upsilon$ ground state, displayed in 
the left and right panel of Fig.~\ref{fig_eps-ups-U-F} when using $U$ 
and $F$, respectively. Note, however, that the scheme dependence of
the binding energy is significantly reduced in the bottomonium case,
to about 10\%, reflecting a better accuracy of the static approximation 
due to the larger bottom-quark mass, as expected. 
With the weaker interaction implicit in $F$ the binding is reduced by 
about a factor of 4. The uncertainty induced by the different lQCD
inputs is significantly larger than the one caused by the reduction
scheme.
%

\section{Influence of the confining force}
\label{app_conf}
%
In this appendix we assess the relevance of the confining interaction 
for bound-state formation.
Recalling the definition of the free energy from Eq.~(\ref{Fcolor-Meg}),
\begin{eqnarray}
F_a(r,T)
&=&V_a^{C}(r,T)+V^{S}(r,T)+2\Sigma_Q(T)\ , 
\nonumber 
\end{eqnarray}
we repeat our calculations with the string term, $V^{S}$, switched off 
while all other parameters are kept fixed. The corresponding results for 
the internal energy, $U$, are obtained using Eq.~(\ref{U}) with the 
modified free energy (we keep, however, the self-energy from the full 
calculation). The results using $U$ as the potential are presented in 
Fig.~\ref{Spec-nostring} using the Thompson reduction. For $J/\psi$ 
($\eta_c$) states the most striking difference occurs in the vacuum 
where without the confining interaction no excited bound states are 
supported and only a modest threshold enhancement remains for the 
ground state. In the medium the relevance of the string term gradually 
decreases until the results become similar to the full calculation for 
a temperature close to 2\,$T_c$ (even though the peak height is still 
smaller). This follows from the significantly stronger screening of the 
confining relative to the Coulomb term ($\tilde m_D(T)$ is much larger 
than $m_D(T)$ above $T_c$, recall Fig.~\ref{fig_para}). Close to $T_c$ 
half of the binding of the $J/\psi$ is still supplied by remnants of 
the confining force. These systematics suggest that charmonia are 
rather sensitive to medium effects on the confining force in the 
temperature regime of 1-2\,$T_c$. At first glance it might surprise 
that the calculation without string term produces more binding in the 
medium than in the vacuum. The reason is that, without the string term, 
the internal energy, as given by Eq.~(\ref{U}), leads to a more 
attractive potential in the medium (up to $\sim$1.5\,$T_c$) than in the 
vacuum (note that we are still using the large effective mass which,
of course, is generated by the large-distance limit of the string term).

For the more tightly bound bottomonia ($\Upsilon$) the sensitivity to 
the string term is still appreciable. In the vacuum the ground state is 
only bound by ca.~100\,MeV while the excited states are unbound. In the 
medium a similar trend as in the charmonium sector is observed, in that 
the significance of the string term ceases as temperature increases.

Generally, our findings clearly demonstrate the importance of the confining 
interaction in both charmonium and bottomonium spectroscopy, both in
vacuum and in medium for temperatures of up to ca.~2\,$T_c$. 
The use of potentials developed in a perturbative expansion therefore 
omits important physics in the description of quarkonium melting in 
medium.

\end{appendix}

\end{document}